%% file: main.tex
\documentclass{article}

\PassOptionsToPackage{sort,numbers}{natbib}
\usepackage[final]{neurips_2022}

\usepackage[utf8]{inputenc} 
\usepackage[T1]{fontenc}    
\usepackage{hyperref}       
\usepackage{url}            
\usepackage{booktabs}       
\usepackage{amsfonts}       
\usepackage{nicefrac}       
\usepackage{microtype}      
\usepackage{xcolor}         

\input{macros}

\title{On the Generalizability and Predictability of Recommender Systems}

\author{Duncan McElfresh%
\thanks{Equal Contribution. Work done while the first two authors were employed at Abacus.AI. Correspondence to: 
\texttt{\{duncan, sujay, colin\}@abacus.ai},
\texttt{valverde@cs.umd.edu},
\texttt{john@arthur.ai}.
}
$^{1}$, Sujay Khandagale$^{*1}$, Jonathan Valverde$^{*1,3}$, \\
\textbf{John P.\ Dickerson$^{2,3}$,  Colin White$^{1}$}
    \vspace*{1mm} \\
    $^1$Abacus.AI, $^2$ArthurAI, $^3$University of Maryland
}

\begin{document}

\maketitle

\input{abstract}

\input{introduction}
\input{relatedwork}

\input{analysis}

\input{reczilla}

\input{conclusion}

\begin{ack}
This work was supported by a GEM Fellowship, NSF CAREER Award IIS-1846237, NIST MSE Award \#20126334, DARPA
GARD \#HR00112020007, and DoD WHS Award
\#HQ003420F0035.
\end{ack}

\bibliography{main}
\bibliographystyle{plain}

\clearpage
\appendix

\input{appendix_experiments}
\input{appendix_reczilla_pipeline}
\input{appendix_results}

\end{document}

%% file: macros.tex
\usepackage{longtable}

\usepackage{adjustbox}
\usepackage{array}

\usepackage{comment}
\usepackage{amsmath,amsthm,amssymb}
\usepackage{algorithm}
\usepackage{algpseudocode}
\usepackage{multicol}
\usepackage{mathtools}
\usepackage{caption}
\usepackage{float}
\usepackage{graphicx}
\usepackage{subcaption}
\usepackage{wrapfig}
\usepackage{multirow}
\usepackage{mathrsfs}
\usepackage{newfloat}
\usepackage{relsize}
\usepackage{enumitem}
\usepackage{pifont}
\usepackage{bm}

\usepackage{lscape}  

\usepackage{rotating}

\newcolumntype{R}[2]{%
    >{\adjustbox{angle=#1,lap=\width-(#2)}\bgroup}%
    l%
    <{\egroup}%
}

\newcolumntype{T}{>$l<$}
\newcolumntype{L}{>{\arraybackslash}m{4cm}}
\newcolumntype{S}{>{\arraybackslash}m{3cm}}

\newcommand*\rot{\multicolumn{1}{R{45}{0.5em}}}

\newcommand{\D}{\mathcal{D}}

\newcommand{\R}{\mathbb{R}}

\usepackage{soul}


%% file: abstract.tex
\begin{abstract}

While other areas of machine learning have seen more and more automation, designing a high-performing recommender system still requires a high level of human effort. Furthermore, recent work has shown that modern recommender system algorithms do not always improve over well-tuned baselines. A natural follow-up question is, ``how do we choose the right algorithm for a new dataset and performance metric?'' In this work, we start by giving the first large-scale study of recommender system approaches by comparing 24 algorithms and 100 sets of hyperparameters across 85 datasets and 315 metrics. We find that the best algorithms and hyperparameters are highly dependent on the dataset and performance metric. However, there is also a strong correlation between the performance of each algorithm and various meta-features of the datasets. Motivated by these findings, we create RecZilla, a meta-learning approach to recommender systems that uses a model to predict the best algorithm and hyperparameters for new, unseen datasets. By using far more meta-training data than prior work, RecZilla is able to substantially reduce the level of human involvement when faced with a new recommender system application. We not only release our code and pretrained RecZilla models, but also all of our raw experimental results, so that practitioners can train a RecZilla model for their desired performance metric: \url{https://github.com/naszilla/reczilla}.
\end{abstract}


%% file: introduction.tex
\section{Introduction} \label{sec:introduction}
Due to the large computational resources for training machine learning models, 
researchers have found many ways to
repurpose existing computation. For example, it is common to start with a pretrained ImageNet classification model for computer vision tasks \citep{deng2009imagenet, ridnik2021imagenet, he2019rethinking}, 
or a pretrained BERT model for natural language tasks \citep{bert, liu2019roberta}.
These approaches work well because the core problems are largely homogeneous; 
for example, any computer vision model
at its core must be able to distinguish edges, colors, and shapes.
Even a task-specific architecture can be found automatically through neural architecture search \citep{nas-survey}, 
since the building blocks such as convolutional layers stay the same.

On the other hand, recommender system (rec-sys) research has followed a different trajectory:
despite their widespread usage across many 
e-commerce, social media, and entertainment companies such as Amazon, YouTube, and Netflix
\citep{covington2016deep, gomez2015netflix, smith2017two},
there is far less work in reusing models or automating the process of selecting models. Many rec-sys techniques are designed and optimized with just a \emph{single} dataset in mind 
\citep{gomez2015netflix, haldar2019applying, covington2016deep, kouki2020lab, wang2019sequential}.
Intuitively, this might be because each rec-sys application is highly unique based on the 
dataset and the target metric.
For example, a typical user session looks very different among e.g.\ YouTube, Home Depot, and AirBnB \citep{covington2016deep, kouki2020lab, haldar2019applying}.
However, this intuition has not been formally established.
Furthermore, recent work has shown that neural recommender system algorithms do not always improve over well-tuned baselines such as $k$-nearest neighbor and matrix factorization \citep{dacrema2021troubling}.
A natural question is then, ``how do we choose the right algorithm for a new dataset and performance metric?''

In this work, we show that the best algorithm and hyperparameters are highly dependent on the dataset and user-defined performance metric.
Specifically, we run the first large-scale study of rec-sys approaches
by comparing 24 algorithms across 85 datasets and 315 metrics.
For each dataset and algorithm pair,
we test up to 100 hyperparameters (given a 10 hour time limit per pair).
The codebase that we release, which includes a unified API for a large, diverse set of algorithms, datasets, and metrics, may be of independent interest.
We show that the algorithms do not \emph{generalize} -- the set of algorithms which perform well changes substantially across dataset and across performance metrics. Furthermore, the best hyperparameters of a rec-sys algorithm on one dataset often perform significantly worse than the best hyperparameters on a different dataset.
Although we show that there are no universal algorithms that work well on most datasets, we \emph{do} show that various meta-features of the dataset can be used
to \emph{predict} the performance of rec-sys algorithms.
In fact, the same meta-features are also predictive of the runtime of rec-sys algorithms as well as the ``dataset hardness'' -- how challenging it is to find a high-performing model on a particular dataset.

\begin{figure}
    \centering
    \includegraphics[width=0.99\textwidth]{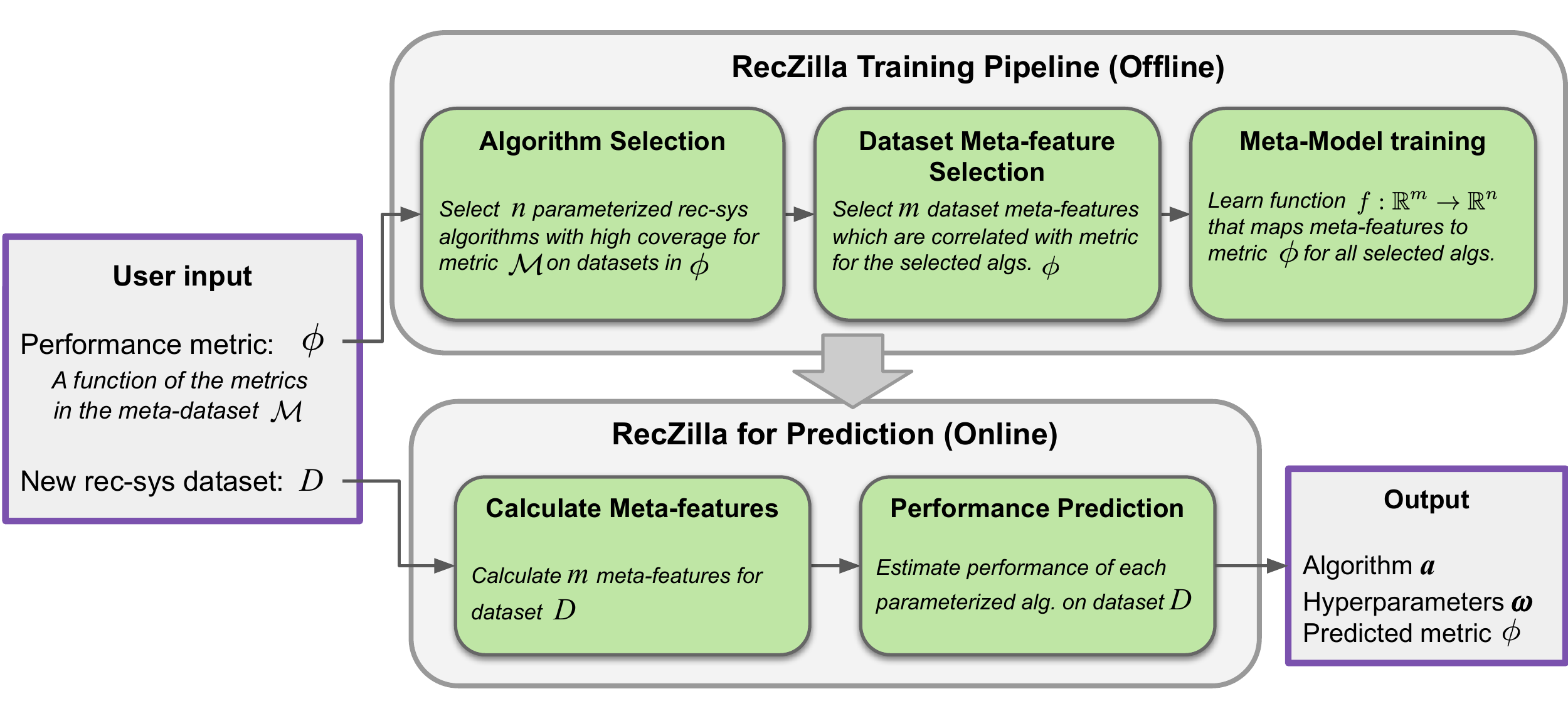}
    \caption{RecZilla recommends a parameterized rec-sys algorithm for a user-provided dataset and performance metric. The RecZilla pipeline is built using a meta-dataset $\mathcal M$ that includes many different performance metrics evaluated on many different rec-sys algorihtms on many different datasets; we estimate algorithm performance using dataset meta-features. To avoid over-fitting and reduce runtime, users can limit the number of algorithms and meta-features to constants $n$ and $m$.}
    \label{fig:reczilla-high-level}
\end{figure}

Motivated by these findings, we introduce RecZilla, a meta-learning-based algorithm selection approach (see Figure \ref{fig:reczilla-high-level}) inspired by SATzilla \citep{satzilla}.
At the core of RecZilla is a model that, given a user-defined performance metric, predicts the best rec-sys algorithm and hyperparameters for a new dataset based on meta dataset features such as number of users and items, and spectral properties of the interaction matrix.
We show that RecZilla quickly finds high-performing 
algorithms on datasets it has never seen before. 
While there has been prior work on 
meta-learning for recommender systems \citep{cunha2018cf4cf, cunha2018metalearning}, no prior work is metric-independent, searches for hyperparameters as well as algorithms, or considers more than nine families of datasets.
By running an ablation study on the number of meta-training datasets, we show that more datasets are crucial to the success of RecZilla.
We release ready-to-use, pretrained RecZilla models for common test metrics, and we release the raw results from our large-scale study, along with code so that practitioners
can easily train a new RecZilla model for their specific performance metric of interest.

\noindent\textbf{Our contributions.}
We summarize our main contributions below.
\begin{itemize}[topsep=0pt, itemsep=2pt, parsep=0pt, leftmargin=5mm]
    \item We run a large-scale study of recommender systems, showing that the best algorithm and hyperparameters are highly dependent on the dataset and user-defined performance metric. We also show that dataset meta-features are predictive of the performance of algorithms.
    \item 
    We create RecZilla, a meta-learning-based algorithm selection approach which, given a performance metric, efficiently predicts the best algorithm and set of hyperparameters on new datasets.
    \item
    We release a public repository containing 85 datasets
    and 24 rec-sys algorithms, accessed through a unified API. Furthermore, we release both pretrained RecZilla models, and raw data so that users can train a new RecZilla model on their desired metric.
\end{itemize}

%% file: relatedwork.tex
\paragraph{Related Work} 

Recommender systems are a widely studied area of research \citep{BOBADILLA2013109}.
Common approaches include $k$-nearest neighbors \citep{adeniyi2016automated}, matrix factorization \citep{koren2009matrix, levy2014neural}, and deep learning approaches
\citep{covington2016deep, gomez2015netflix, smith2017two}. For a survey on recommender systems, see \citep{BOBADILLA2013109, aggarwal2016recommender}.
A recent meta-study showed that of the 12 published neural rec-sys approaches published at top conferences between 2015 and 2018, 11 performed worse than well-tuned baselines (e.g.\ nearest neighbor search or linear models) \citep{dacrema2021troubling}. 
Another recent paper found that the relative performance of rec-sys algorithms can change significantly based on the choice of datasets used \citep{chin2022datasets}.

Algorithm selection for recommender systems was first studied in 2011 \citep{huang2011does}
by using a graph representation of item ratings.
Follow-up work used dataset meta-features to select the best nearest neighbor and matrix factorization
algorithms \citep{ekstrand2012recommenders, adomavicius2012impact, griffith2012investigations}.
Subsequent work focused on improving the model and framework \citep{cunha2018metalearning}
including studying 74 meta-features systematically \citep{cunha2016selecting}.
More recent approaches from 2018 run meta-learning for recommender systems by casting the meta-problem
itself as a collaborative filtering problem. Performance is then estimated with 
subsampling landmarkers \citep{cunha2017recommending, cunha2018cf4cf, cunha2018cf4cfmeta}.
No prior work in algorithm selection for rec-sys includes open-source Python code. 
There is also work on automated machine learning (AutoML) for recommender
systems, without meta-learning \citep{auto-rec, auto-surprise, gupta2021enpso, gupta2020auto}.
%
Finally, we note that meta-learning across rec-sys datasets is not to be confused with the body of work on meta-learning user preferences \emph{within a single} rec-sys dataset 
\citep{collins2019meta, collins2018novel, collins2018one}.
To the best of our knowledge, no meta-learning or AutoML rec-sys paper has run experiments on more 
than nine dataset families or four test metrics, and no prior work predicts hyperparameters in addition to algorithms.

%% file: analysis.tex
\section{Analysis of Recommender Systems} \label{sec:analysis}

In this section, we present a large-scale empirical study of rec-sys algorithms across a large, diverse set of datasets and metrics.
We assess the following two research questions.
\begin{enumerate}[topsep=0pt, itemsep=2pt, parsep=0pt, leftmargin=5mm]
    \item \textbf{Generalizability.} If a rec-sys algorithm or set of hyperparameters performs well on one dataset and metric, will it perform well on other datasets or on other metrics?
    \item \textbf{Predictability.} Given a metric, can various dataset meta-features be used to predict the performance of rec-sys algorithms?
\end{enumerate}

\paragraph{Algorithms, datasets, and metrics implemented.}
We present full results for 20 rec-sys algorithms, including methods from recent literature and common baselines. Methods include five similarity and clustering-based methods: User-KNN~\cite{userknn}, Item-KNN~\cite{itemknn}, P3Alpha~\cite{cooper2014random}, RP3Beta~\cite{paudel2016updatable}, and Co-Clustering~\cite{george2005scalable}; six Matrix-Factorization (MF) methods: MF-FunkSVD, MF-AsySVD~\cite{asysvd}, MF-BPR~\cite{rendle2012bpr}, IALS~\cite{hu2008collaborative}, Pure-SVD, Non-negative matrix factorization (NMF)~\cite{cremonesi2010performance}; five methods based on linear models: Global-Effects, SLIM-BPR~\cite{slim-bpr}, SLIM-Elastic-Net~\cite{levy2013efficient}, EASE-R~\cite{steck2019embarrassingly}, and SlopeOne~\cite{lemire2005slope}; 
two simple baselines: Random and Top-Pop; and two neural network based methods: User-NeuRec~\cite{neurec} and Mult-VAE~\cite{multi-vae}.
We also include partial results (on 8-10 datasets each) for four more neural network based methods: 
DELF-EF \citep{ijcai2018-462}, 
DELF-MLP \citep{ijcai2018-462}, 
Item-NeuRec \citep{neurec}, 
and Spectral-CF \citep{spectral-cf} (with results included in Tables \ref{tab:all-alg-ranks}, \ref{tab:all-alg-ranks-b}, and \ref{tab:all-alg-ranks-c}).
These algorithms were chosen due to their high performance, popularity, and speed.
For many algorithms, we used the implementations from the codebase of Dacrema et al.\ \citep{dacrema2021troubling}.
For full details of the algorithms, see Appendix \ref{app:experiments}.

We run the algorithms on 85 datasets from 19 dataset ``families'':
Amazon \citep{ni-etal-2019-justifying}, Anime \citep{cho2018analyzing},
BookCrossing \citep{book-crossing}, CiaoDVD \citep{konect,etaf-ciaodvd}, Dating (Libimseti.cz) \cite{konect,online-dating}, Epinions \citep{massa2008trustlet,trust-aware-recsys}, FilmTrust \citep{guo2013novel}, Frappe \cite{frappe15}, Gowalla \citep{gowalla}, Jester2 \citep{jester2}, LastFM \cite{Cantador:RecSys2011}, MarketBias-Electronics and MarketBias-ModCloth \cite{wan-marketbias}, MovieTweetings \cite{Dooms13crowdrec}, Movielens \cite{movielens}, NetflixPrize \citep{netflixprize},
Recipes \cite{majumder-etal-2019-generating}, Wikilens \cite{frankowski-wikilens}, and Yahoo \citep{dror2012yahoo}.
%
Here, a ``dataset family'' refers to an original dataset, while ``dataset'' refers to a single train-test split drawn from the original dataset, which may be a small subset of the original.
We implemented the majority of rec-sys datasets commonly used for research; to the best of our knowledge, this is the largest number of rec-sys datasets accessible in a single open-source repository.

We use 23 different ``base'' metrics: 
ARHR,
Average Popularity,
Diversity Similarity,
F1 Score,
Gini Index, 
Herfindahl Index, 
Hit Rate,
Item Coverage,
Item-hit Coverage,
Precision (PREC), 
Precision-Recall Min Denominator,
Recall, 
MAP, 
MAP Min Denominator,
Mean Inter-List Diversity, 
MRR, 
NDCG,   
Novelty, 
Shannon Entropy,
User Coverage, 
User-hit Coverage.
All of these metrics are computed at cutoffs $\{1, 2, 3, 4, 5, 6, 7, 8, 9, 10, 15, 20, 30, 40, 50\}$, for a total of 315 different metrics. See Appendix \ref{app:metrics} for more details.
These metrics include the most popular from the literature, and we use the Dacrema et al.\ \citep{dacrema2021troubling} implementations.
%

\input{tables/table_3_new}

\paragraph{Experimental design.}
Each dataset's train, validation, and test split is based on leave-last-$k$-out (and our repository also includes splits based on global timestamp).
We use a random hyperparameter search for all methods, with the exception of neural network based methods. Since neural networks require far more resources to train (longer training time, and requiring GPUs), we use only the default hyperparameters for neural algorithms. 
For each non-neural algorithm, we expose several hyperparameters and give ranges based on common values. For each dataset, we run each algorithm on a random sample of up to 100 hyperparameter sets. Each algorithm is allocated a 10 hour limit for each dataset split; we train and test the algorithm with at most 100 hyperparameter sets on an \texttt{n1-highmem-2} Google Cloud instance, until the time limit is reached.
Each neural network method is trained on each dataset using the default hyperparameters used in its respective paper, with a time limit of 15 hours on 
an NVIDIA Tesla T4 GPU. All neural network methods are trained with batch size 64, for up to 100 epochs; early stopping occurs if loss does not improve in 5 epochs.

Each algorithm is trained on the train split, and the performance metrics are computed on the test split.
We refer to each combination of (algorithm, hyperparameter set, dataset) as an \emph{experiment}.
By running 24 algorithms, most with up to 100 hyperparameters, on 85 datasets, this resulted in 84\,850 successful experiments, and by computing 315 metrics, our final meta-dataset of results includes more than 26 million evaluations.
We give a more detailed look at the breakdown of experiments in Appendix \ref{app:experiments}, and we discuss any potential biases in the resulting dataset in Section \ref{sec:conclusion}.

\subsection{On the generalizability of rec-sys algorithms} \label{subsec:generalizability}

\emph{If a rec-sys algorithm or set of hyperparameters performs well on one dataset and metric, will it perform well on other datasets or on other metrics?}

Our first observation is that \emph{all} algorithms perform \emph{well} on some datasets, and poorly on others.
First we identify the best-performing hyperparameter set for each (algorithm, dataset) pair---to simulate hyperparameter optimization using our meta-dataset.
We then rank all algorithms for each dataset, according to several performance metrics. 

If we focus on a single metric, then many algorithms are ranked first according to this metric on at least one dataset.
Take for example metric NDCG@1: 17 of 20 algorithms are ranked either first or second on at least one dataset. The same is true for metric RECALL@50: all algorithms except for SlopeOne, GlobalEffects, and Random are ranked either first or second on at least one dataset. The same is true for many other metrics and cutoffs (see Table~\ref{tab:all-alg-ranks} in Appendix~\ref{app:results}).

Average performance is more varied: some algorithms tend to perform better than others. 
Table~\ref{tab:overall-ranks} shows the mean, min (best) and max (worst) ranking of all 24 algorithms over all dataset and all accuracy and hit-rate metrics. This includes metrics NDCG, precision, recall, Prec.-Rec.-Min-density, hit-rate, F1, MAP, MAP-Min-density, ARHR, and MRR (see Appendix~\ref{app:metrics} for descriptions of these metrics).
Nearly all algorithms are ranked first for at least one metric on at least one dataset.
Many algorithms perform well on average. 
Furthermore, most algorithms perform very poorly in some cases: the maximum rank is at least 14 (out of 20) for all algorithms.

\begin{figure}
    \centering
    \includegraphics[width=0.99\textwidth]{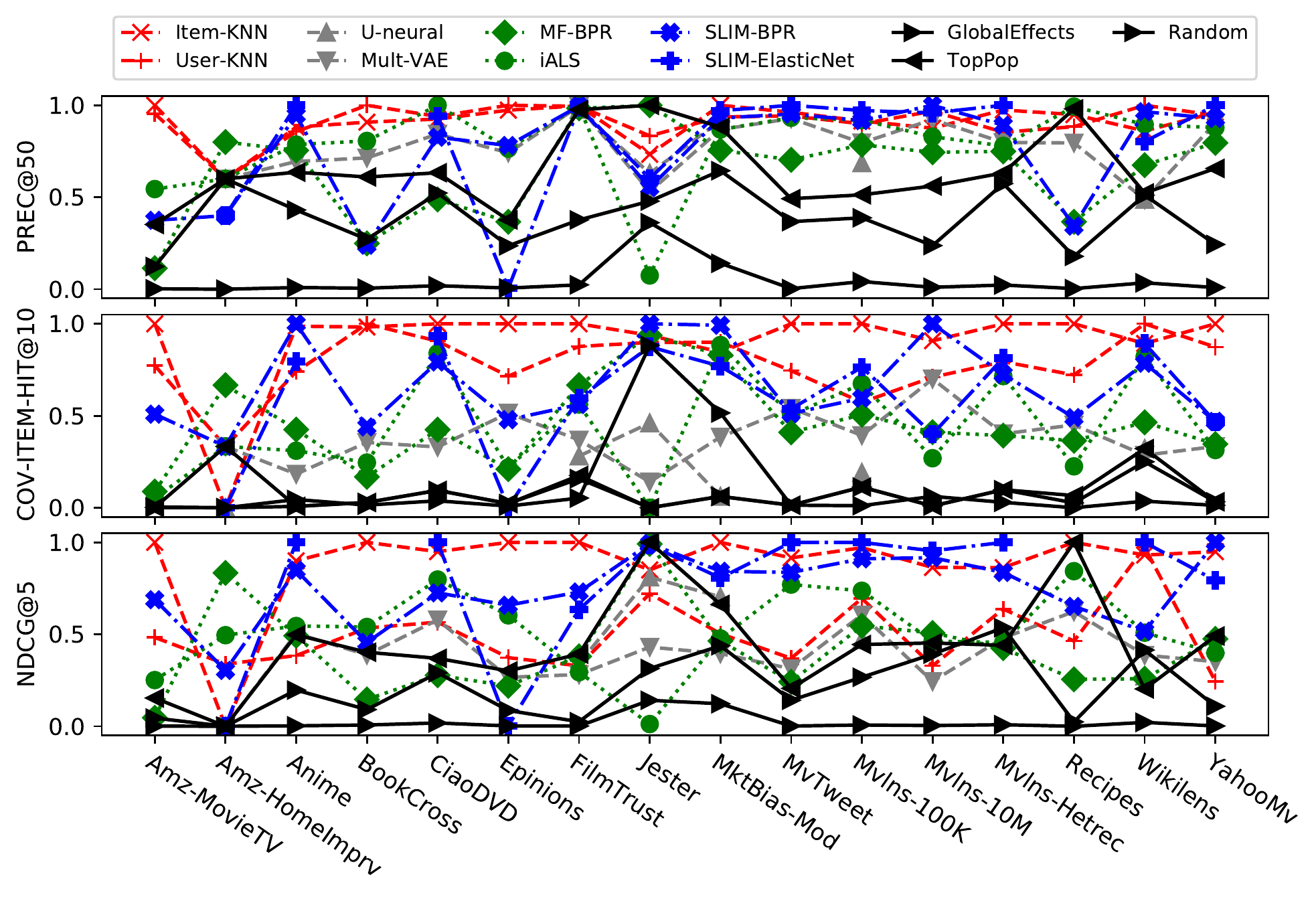}
    \caption{Relative algorithm performance depends on both the dataset and metric: no algorithm dominates across all dataests or metrics. Each plot shows a different metric, normalized to $[0, 1]$ for each dataset; the horizontal axis shows different dataset, ordered alphabetically. Each series corresponds to a different algorithm: similarity-based methods are red, matrix factorization methods are green, baseline methods are black, and neural network methods are in gray.}
    \label{fig:squiggly}
\end{figure}

To illustrate the changes in algorithm performance across datasets, Figure~\ref{fig:squiggly} shows the normalized metric values for eight algorithms across 17 dataset splits.
Some algorithms tend to perform well (Item-KNN and SLIM-BPR) and others poorly (Random, TopPop), but no algorithm clearly dominates for all metrics and datasets.
This is a primary motivation for our meta-learning pipeline descirbed in Section~\ref{sec:reczilla}:  different algorithms perform well for different datasets on different metrics, so it is important to identify appropriate algorithms for each setting.

\paragraph{Generalizability of hyperparameters.}
While the previous section assessed the generalizability of pairs of (algorithm, hyperparameters), now we assess the generalizability of the hyperparameters themselves while keeping the algorithms fixed.
For a given rec-sys algorithm, we can tune it on a dataset $i$, and then evaluate the normalized performance of the tuned method on a dataset $j$, compared to the normalized performance of the best hyperparameters from dataset $j$. In other words, we compute the performance of tuning a method on one dataset and deploying it on another.

\begin{figure}
     \centering
     \begin{subfigure}[b]{0.48\textwidth}
         \centering
         \includegraphics[width=\textwidth]{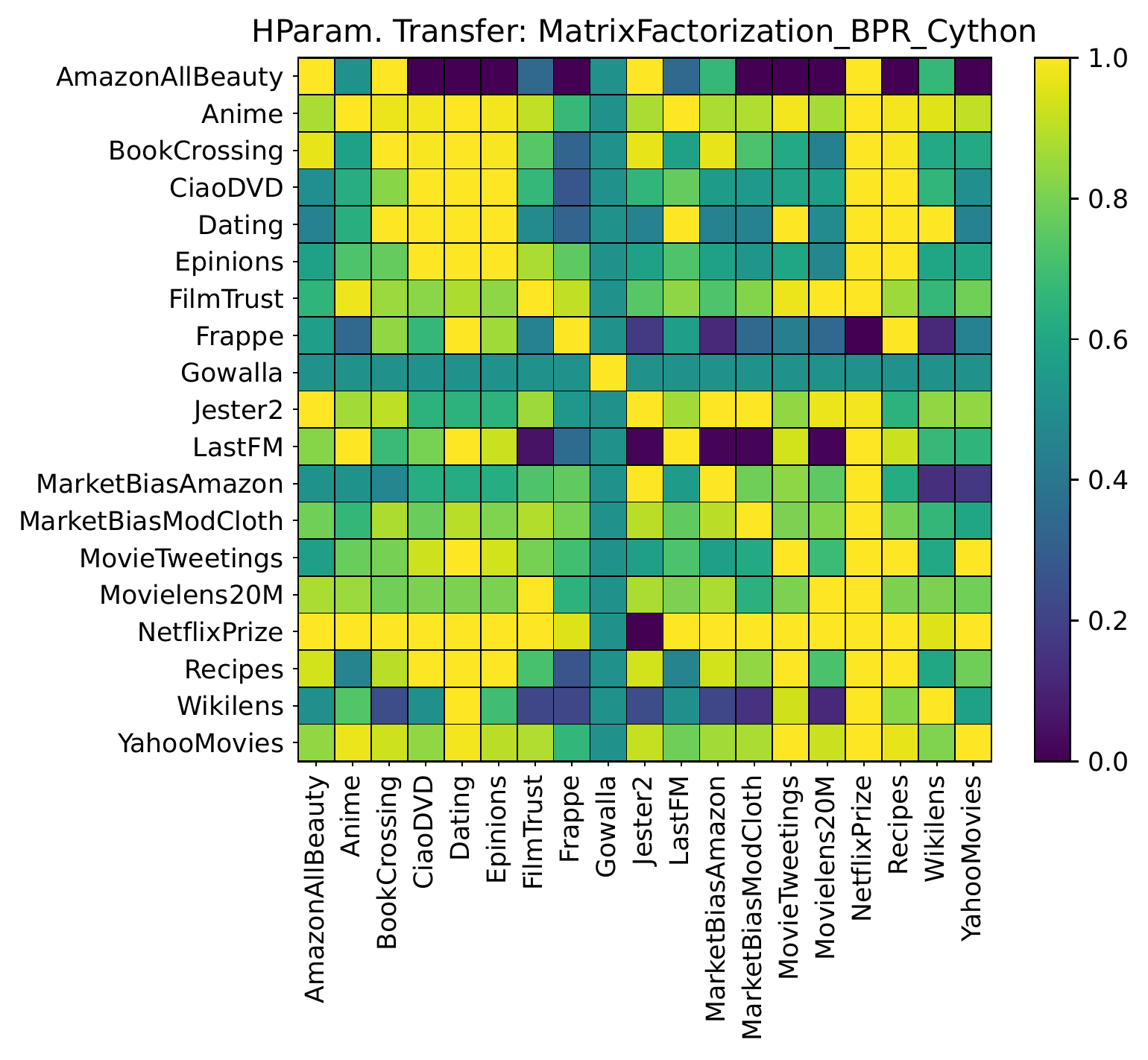}
     \end{subfigure}
     \hfill
     \begin{subfigure}[b]{0.48\textwidth}
         \centering
         \includegraphics[width=\textwidth]{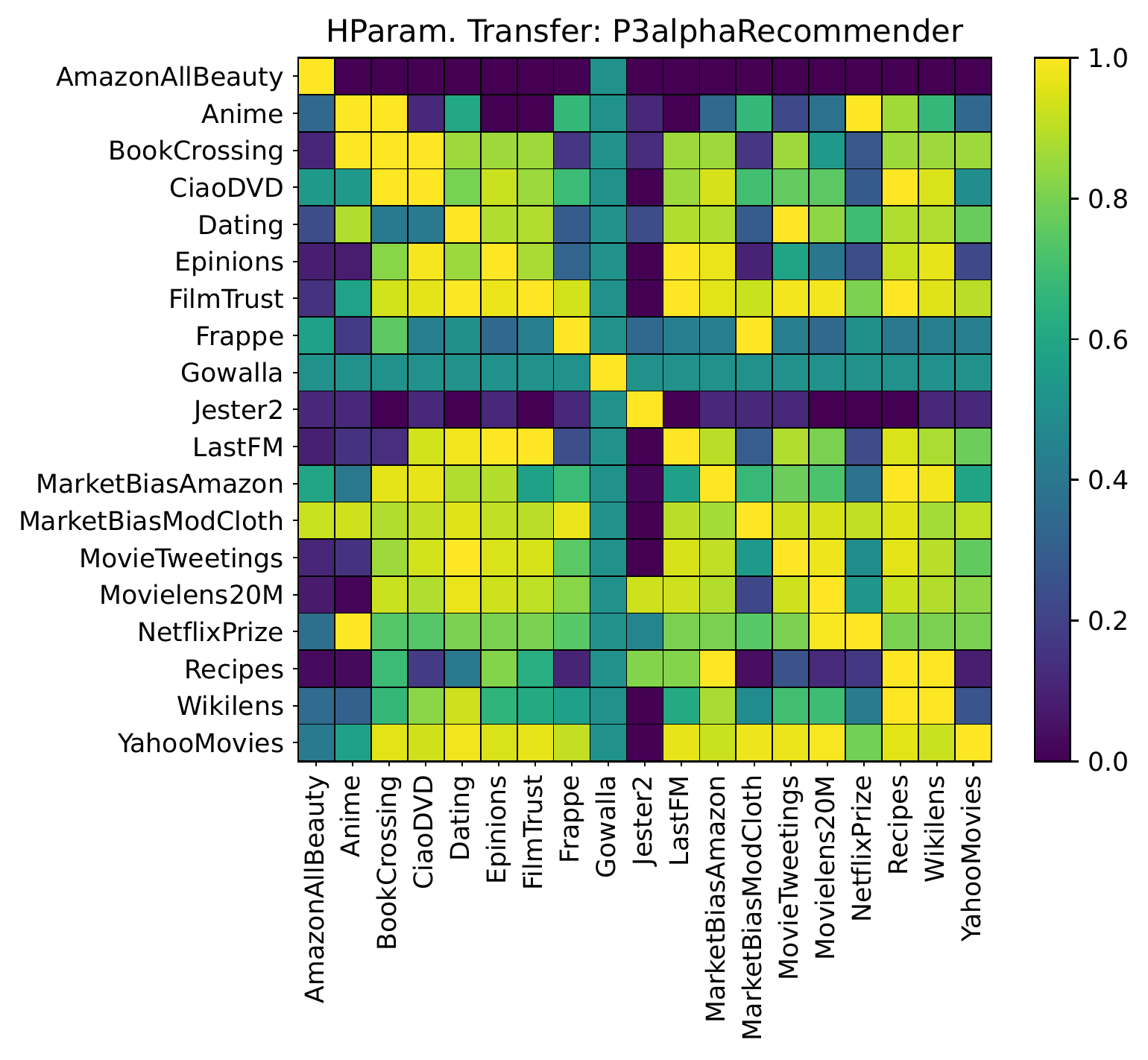}
     \end{subfigure}
     \centering
     \begin{subfigure}[b]{0.48\textwidth}
         \centering
         \includegraphics[width=\textwidth]{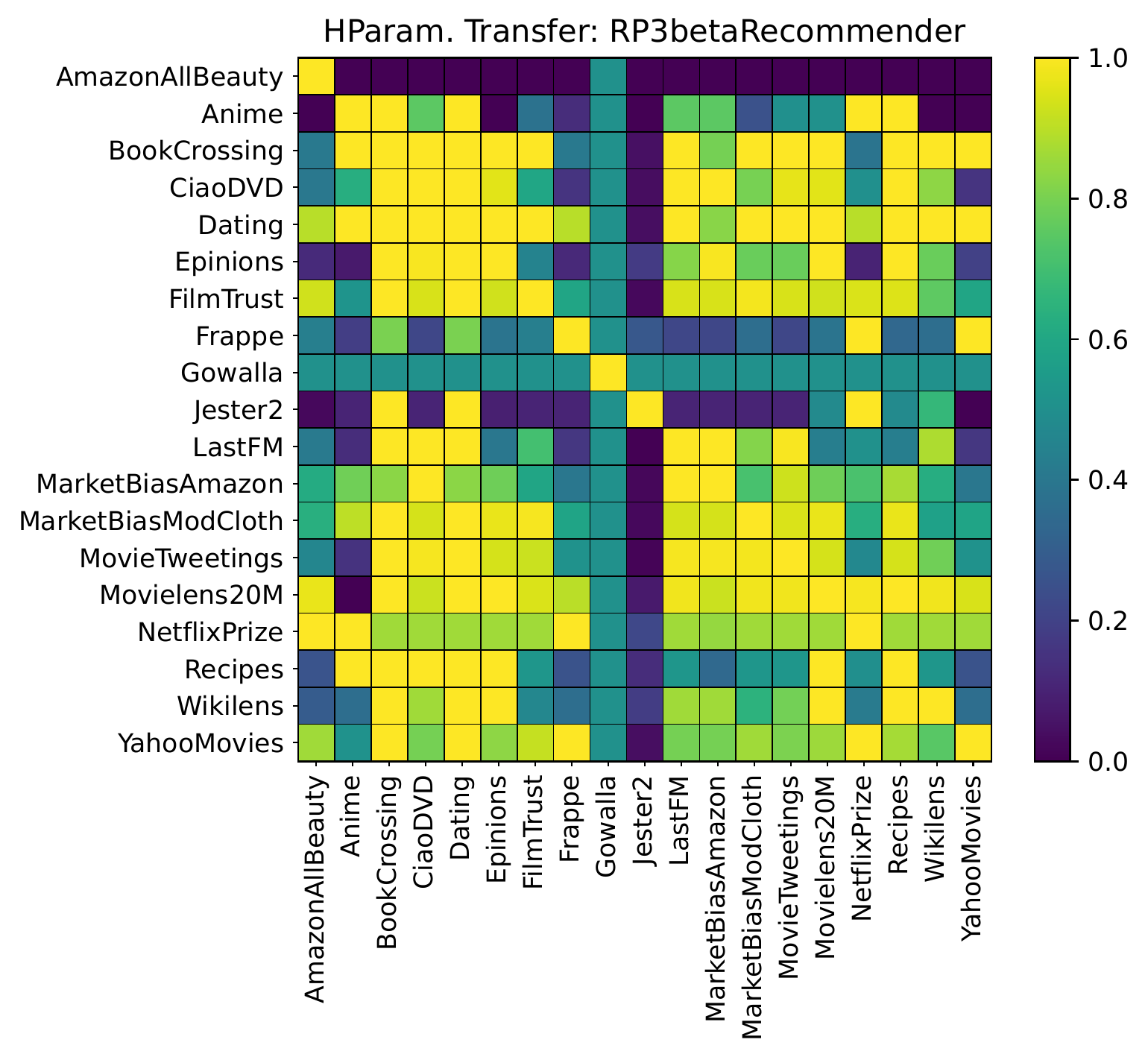}
     \end{subfigure}
     \hfill
     \begin{subfigure}[b]{0.48\textwidth}
         \centering
         \includegraphics[width=\textwidth]{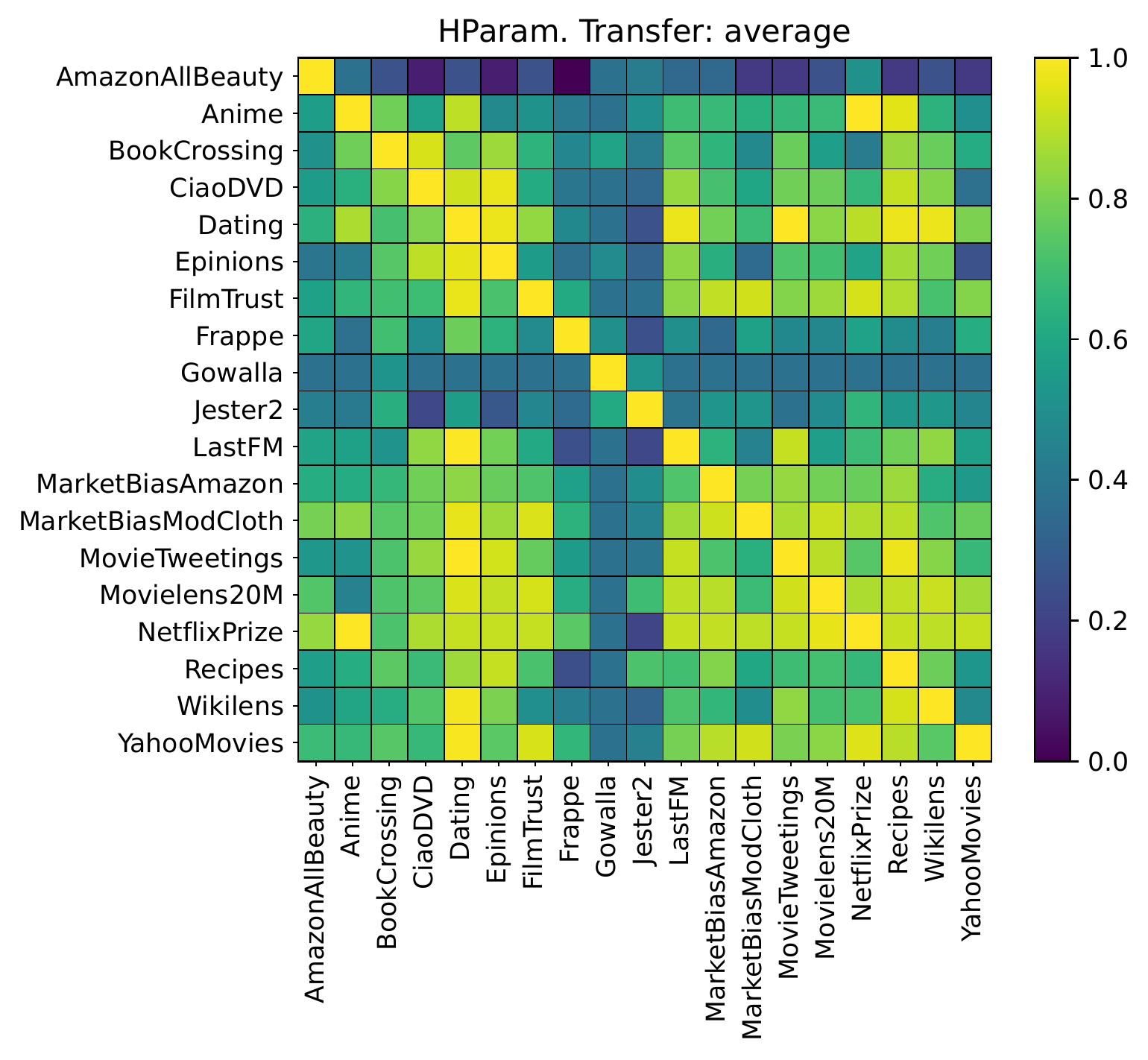}
     \end{subfigure}
        \caption{Transferability of hyperparameters across datasets, for three different algorithms, and the average of all algorithms (bottom right).
        For each plot, row $i$, column $j$ denotes the relative performance of an algorithm tuned on dataset $i$ and then evaluated on dataset $j$.
        A value close to 1 indicates that the hyperparameters transfer well from $i$ to $j$, while a value close to 0 indicates that the hyperparameters transfer poorly.
        }
        \label{fig:matrices}
\end{figure}

In Figure \ref{fig:matrices}, we run this experiment for all pairs of datasets (one dataset per dataset family). We plot the hyperparameter transfer for three different algorithms, as well as the average over all algorithms which completed sufficiently many experiments across the set of hyperparameters.
For each given $i$, $j$, we create the set of hyperparameters that completed for the given algorithm on both datasets $i$ and $j$, and then we use min-max scaling for the performance metric values of these hyperparameters on $i$ and on $j$ separately. Therefore, all matrix values are between 0 and 1; a value close to 1 indicates that the best hyperparameters from dataset $i$ are also nearly the best on dataset $j$. A value close to 0 indicates that the best hyperparameters from dataset $i$ are nearly the worst for dataset $j$.
Across all algorithms, we see that it is particularly hard for hyperparameters to transfer to and from some datasets such as Gowalla and Jester2.
Furthermore, the majority of pairs of datasets do not have strong hyperparameter transfer.
Overall, these experiments give evidence that tuning the hyperparameters of an algorithm on one dataset and transferring to another dataset does not give high performance, motivating RecZilla which predicts the best hyperparameters for a given algorithm and dataset.

\subsection{On the predictability of rec-sys algorithms} \label{subsec:predictability}

\emph{Can attributes of the rec-sys dataset be used to predict the performance of rec-sys algorithms?}

\paragraph{Dataset meta-features.}
We calculate 383 different meta-features to characterize each dataset.
These meta-features include statistics on the rating matrix---including basic statistics, the distribution-based features of Cunha et al.\ \citep{cunha2016selecting}, and landmark features~\citep{cunha2017recommending}---which measure the performance of simple rec-sys algorithms on a subset of the training dataset.
Since these meta-features are used for algorithm selection, they are calculated using only the training split of each dataset. 
For more details on the dataset meta-features, see Appendix \ref{app:meta-features}.

\paragraph{Algorithm performance prediction.}

\begin{table}[t]
\caption{Highest absolute correlations computed across 85 datasets and weighed inversely proportional to dataset family frequency, over all pairs of algorithm families and meta-features, for the PREC@10 performance metric and the default algorithm hyperparameters.}
\centering
\begin{tabular}{@{}l|l|l@{}}
\toprule
\multicolumn{1}{l}{\textbf{Abs. Correlation}} & \multicolumn{1}{l}{\textbf{Algorithm Family}} & \multicolumn{1}{l}{\textbf{Meta-feature}} \\
\midrule 
0.941                & SlopeOne                  & Mean of item rating count distribution   \\
0.933                & CoClustering              & Median of item rating count distribution \\
0.887                & MF-BPR                    & Sparsity of rating matrix                \\
0.855                & RP3beta                   & Mean of item rating count distribution   \\
0.846                & UserKNN                   & Landmarker, Pure SVD, mAP@5  \\
\bottomrule
\end{tabular}
\label{tab:top-corrs}
\end{table}

Table \ref{tab:top-corrs} shows the meta-features that are most highly-correlated with the performance (PREC@10) of each algorithm, using their default parameters. 
Several meta-features aare highly-correlated  with algorithm performance; one of the simplest metrics---the mean of the item rating count distribution---is highly correlated with performance of two rec-sys algorithms.
This experiment motivates the design of RecZilla in the next section, which trains a model using dataset meta-features to predict the performance of algorithms on new datasets.

As a toy-model version of RecZilla, we train three different meta-learner functions (XGBoost, KNN, and linear regression) using our meta-dataset, to predict performance metric PREC@10 for 10 rec-sys algorithms with high average performance (see Appendix~\ref{app:experiments} for details). 
We use leave-one-out evaluation for each meta-learner: one dataset family is held out for testing, while $m$ are used for training. 
Figure~\ref{fig:meta-learner} shows the mean absolute error (MAE) of each meta-learner; these results are aggregated over 200 random samples of randomly-selected training dataset families.
MAE decreases as more dataset families are added, suggesting that it is possible to estimate rec-sys algorithm performance using dataset meta-features.

We also find that performance metrics are not the only values that can be predicted with dataset meta-features.
In particular, we find that the \emph{runtime} of rec-sys algorithms is also highly correlated with different meta-features.
Furthermore, we compute a simple measure of \emph{dataset hardness}, which we compute as, given a performance metric, the maximum value achieved for that dataset across all algorithms. For example, if all 20 algorithms do not perform well on the MovieTweetings dataset, then we can expect that the MovieTweetings dataset is ``hard''.
Once again, we find that certain dataset meta-features are highly correlated with dataset hardness.
For more details on meta-feature correlation with algorithm runtimes and dataset hardness, see Appendix \ref{app:experiments}.

The fact that dataset meta-features are correlated with algorithm performance, algorithm runtimes, and dataset hardness is a strong positive signal that meta-learning is worthwhile and useful in the context of recommender systems. 
We explore this direction further in the next section.

%% file: tables/table_3_new.tex
\begin{table}
\caption{The relative performance of each rec-sys algorithm depends on the dataset and metric. This table shows the mean, min (best) and max (worst) rank achieved by all 20 algorithms over all 85 datasets, over 10 accuracy and hit-rate metrics at all cutoffs tested. This includes metrics NDCG, precision, recall, Prec.-Rec.-Min-density, hit-rate, F1, MAP, MAP-Min-density, ARHR, and MRR. \label{tab:overall-ranks}}
\tabcolsep3pt 
\centering
\resizebox{\columnwidth}{!}{%
\begin{tabular}{lllllllllllllllllllll}
\toprule
Rank & \rot{Item-KNN} & \rot{P3alpha} & \rot{SLIM-BPR} & \rot{EASE-R} & \rot{RP3beta} & \rot{SVD} & \rot{SLIM-ElasticNet} & \rot{iALS} & \rot{NMF} & \rot{User-KNN} & \rot{MF-Funk} & \rot{TopPop} & \rot{MF-Asy} & \rot{MF-BPR} & \rot{Mult-VAE} & \rot{U-neural} & \rot{GlobalEffects} & \rot{CoClustering} & \rot{Random} & \rot{SlopeOne} \\
\midrule
Min.  &              1 &             1 &              1 &            1 &             1 &         1 &                     1 &          1 &         1 &              1 &             1 &            1 &            1 &            1 &              1 &              1 &                   2 &                  1 &            9 &              7 \\
Max.  &             14 &            18 &             14 &           18 &            17 &        16 &                    17 &         19 &        14 &             17 &            18 &           19 &           16 &           17 &             20 &             20 &                  20 &                 19 &           20 &             20 \\
Mean  &            2.3 &           4.2 &            4.7 &          5.3 &             6 &         6 &                     7 &          7 &       7.1 &            7.6 &           9.4 &         10.4 &         10.7 &         11.2 &           11.7 &           12.3 &                13.3 &               14.9 &         16.2 &           16.7 \\
\bottomrule
\end{tabular}

}
\end{table}


%% file: reczilla.tex
\section{RecZilla: Automated Algorithm Selection} \label{sec:reczilla}

In the previous section, we found that \emph{(1)} the best algorithm and hyperparameters strongly depend on the dataset and user-chosen performance metric, and \emph{(2)} the performance of algorithms can be predicted from dataset meta-features.
Points \emph{(1)} and \emph{(2)} naturally motivate an algorithm selection approach to rec-sys powered by meta-learning.

In this section, we present \emph{RecZilla}, which is motivated by a practical challenge: given a performance metric and a new rec-sys dataset, quickly identify an algorithm and hyperparameters that perform well on this dataset. 
This challenge arises in many settings, e.g., when selecting good baseline algorithms for academic research, or when developing high-performing rec-sys algorithms for a commercial application. We begin with an overview and then formally present our approach.

\paragraph{Overview.}
\emph{RecZilla} is an algorithm selection approach powered by meta-learning.
We use the results from the previous section as the meta-training dataset.
Given a user-specified performance metric, we train a meta-model that predicts the 
performance of each of a set of algorithms and hyperparameters on a dataset, 
by using meta-features of the dataset.
Given a new, unseen dataset, we compute the meta-features of the dataset, and then use the meta-model to predict the performance of each algorithm, returning the best algorithm according to the performance metric.
See Figure \ref{fig:reczilla-high-level}.

\begin{wrapfigure}{r}{0.5\textwidth}
\begin{center}
    \centering
\includegraphics[width=0.49\textwidth]{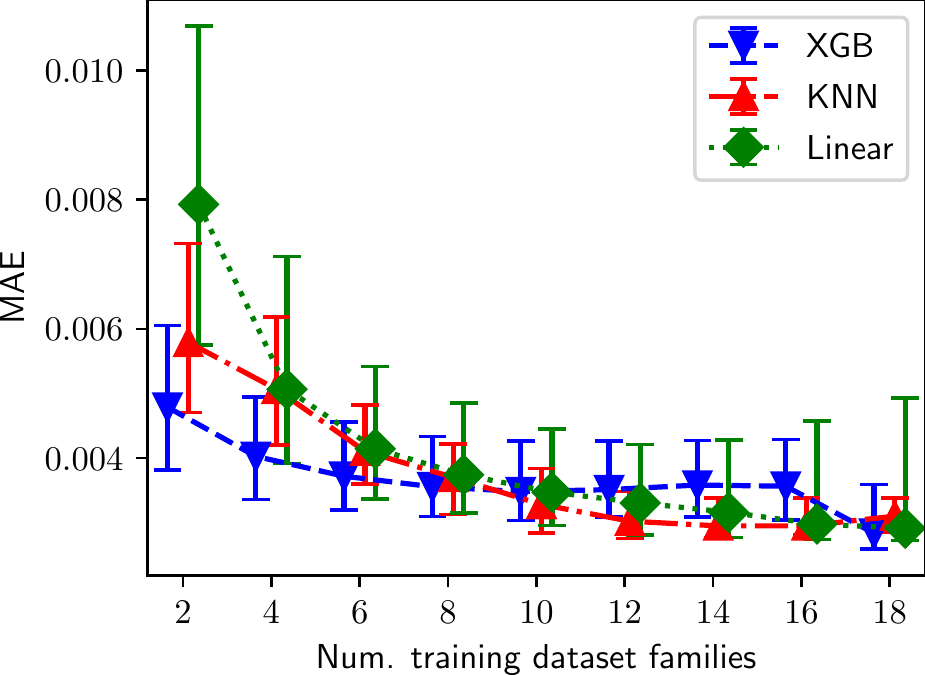}
\end{center}
\caption{Three basic meta-learners (KNN, linear regression, and XGB) are trained randomly-selected dataset families to predict performance metric PREC@10. As more dataset families are added, the meta-learners are better able to predict rec-sys algorithm performance, suggesting that our dataset meta-features are useful for predicting rec-sys algorithm performance. Vertical axis shows mean absolute error (MAE), over all folds of leave-one-out validation, and 200 random trials; in each trial a different set of training datasets are chosen. Error bars show the 40th and 60th percentile.}		\label{fig:meta-learner}
\vspace{-5mm}
\end{wrapfigure}

\paragraph{Preliminaries.}
We start with notation for the rec-sys problem.
Let $D$ denote a rec-sys dataset, consisting of a set of user-item interactions,
split into a training and validation set. 
Let $a$ denote a rec-sys algorithm parameterized by a set of $k({a})$ hyperparameters 
$\bm \omega \in H(a)\subseteq \R^{k(a)}$, which is algorithm-dependent.
Suppose we train algorithm $a$ on the training split of dataset $D$, using hyperparameters $\bm \omega$; 
we denote the \emph{performance} of algorithm $a$ with hyperparameters $\bm \omega$ on dataset $D$ as $\phi(a, \bm \omega, D)\in \R$. The function $\phi(\cdot, \cdot, \cdot)$ 
represents a \emph{performance metric} for the recommender system that is selected by the user; throughout this paper we refer to this function and its numerical value simply as \emph{performance}.
In this paper, larger values of $\phi(\cdot, \cdot, \cdot)$ indicates better performance.
%
We report the \emph{normalized} performance defined as 
\begin{equation*}
\overline{\phi}(a, \bm \omega, D) \equiv 100\times \frac{ \phi(a, \bm \omega, D) - P_D^{min}}{P_D^{max} - P_D^{min}},
\end{equation*}
where $P_D^{max}$ (resp.\ $P_D^{min}$) are the max (resp.\ min) performance of any algorithm on $D$.

Next we define notation for the meta-learning problem.
Given a fixed performance function $\phi$, let $\mathcal{M} = \left\{(D_i, a_i, \bm \omega_i, y_i)\right\}_{i=1}^M$ denote a \emph{meta-dataset} consisting of $M$ tuples. 
%
Each tuple consists of a dataset $D$, algorithm $a$ parameterized by $\bm \omega$, and performance $y$, where $y_i\equiv \phi(a_i, \bm \omega_i, D_i)$.
Since many algorithms have a wide range of hyperparameters, we refer to the set of \emph{parameterized algorithms} in the meta-dataset as $\mathcal S = \{(a_i, \bm \omega_i)\}_{i=1}^N$.
We represent each dataset using vector $\bm d\in \R^m$, where each element of $\bm d$ is a meta-feature of the dataset. 
In this setting, the meta-learning task is to identify a function that estimates the performance of parameterized algorithm $(a, \bm \omega)$ on dataset $D$.

\subsection{The RecZilla Algorithm Selection Pipeline}\label{sec:reczilla-pipeline}

The RecZilla algorithm selection pipeline takes as input a meta-dataset $\mathcal{M}$ built with a user-chosen performance function for the recommendation task $\phi(\cdot, \cdot, \cdot)$.
RecZilla can accommodate any performance function that is a computable function of the performance metrics present in the meta-dataset.
The RecZilla pipeline proposed here returns both an algorithm $a\in \mathcal A$ \emph{and} a set of hyperparameters $\bm \omega \in H(a)$, so that RecZilla can be used without additional hyperparameter optimization.

Since the meta-learning problem is in the low-data regime,
to guard against overfitting, we select a subset of the algorithms that have good coverage over the
training dataset. We similarly select only the meta-features which are most predictive of the user-selected performance metric for the selected algorithms.
Below we outline each of the steps used to build the proposed RecZilla pipeline.

\begin{enumerate}[topsep=0pt, itemsep=2pt, parsep=0pt, leftmargin=5mm]
    \item \textbf{Algorithm subset selection:} 
    We select a subset $\mathcal{S}'\subseteq \mathcal{S}$ of $n$ parameterized algorithms that collectively perform
    well on all datasets in the meta-training dataset $\mathcal{M}$, according to performance function $\phi$.
    %
    We aim to select an algorithm subset with high \emph{coverage} over the set of known datasets, where coverage of a subset $\mathcal{S}'$ is defined as
    \begin{equation} \label{eq:algo-coverage}
    C(\mathcal{S}') = \frac{1}{|\D|}\sum\limits_{D\in \D}
    %
    \max_{(a, \bm \omega) \in \mathcal S'}\overline{\phi}(a, \bm \omega, D).
    \end{equation}
    In other words, the coverage of subset $\mathcal S'$ is the normalized performance metric of the \emph{best} performing algorithm in $\mathcal S'$, averaged over all datasets $\mathcal D$.
    Selecting a subset with maximum coverage is itself a difficult problem; we use a greedy heuristic as follows.
    We begin with $\mathcal S'=\{\}$ and iteratively add the parameterized algorithm $(a^*, \bm \omega^*)\in \arg\max_{(a, \bm \omega)\in \mathcal S} C(\mathcal S' \cup \{(a, \bm \omega)\})$ to $\mathcal{S}'$ until $|\mathcal{S}'|=n$. That is, we greedily ensure that the coverage is maximized at each step.
    \item \textbf{Meta-feature selection:} 
    Similar to the previous point, we select a subset of meta-features with good coverage over the meta-training dataset $\mathcal{M}$, where here coverage is defined in terms of the correlation between algorithm performance and each meta-feature (see Appendix~\ref{app:meta-features} for details).     %
    Let $M: \mathcal D \rightarrow \mathbb R^m$ denote the resulting function that maps a dataset to a vector of meta-features.
    \item \textbf{Meta-learning:} We learn a function $f: \mathbb R^m \rightarrow \mathbb R^n$, where $f(\bm d)=\hat{\bm y}$ is a vector of the estimated performance metric of all parameterized algorithms in $\mathcal S$ on dataset meta-features $\bm d$.
\end{enumerate}

Note that the RecZilla pipeline has two hyperparameters: $n$, the number of parameterized algorithms in $\mathcal S'$; and $m$, the number of dataset meta-features used by the meta-learner.
In our experiments, we run an ablation study with both $n$ and $m$, as well as different functions $f$ for the meta-learning model.

\paragraph{Using RecZilla for Algorithm Selection.}
After developing the RecZilla pipeline, we use the following steps to select an algorithm for a new dataset $D'$:
\begin{enumerate}[topsep=0pt, itemsep=2pt, parsep=0pt, leftmargin=5mm]
    \item Calculate $m$ meta-features of the dataset $\bm d' \gets M(D')$.
    \item Estimate the performance of all parameterized algorithms: $\bm y' \gets f(\bm d')$.
    \item Return the parameterized algorithm in $\mathcal S'$ with the best estimated performance. 
\end{enumerate}

\subsection{Experiments}\label{subsec:experiments}

In this section, we evaluate the end-to-end RecZilla pipeline.
We start by describing the specific versions of RecZilla used in our experiments.
We use four different meta-learning functions within RecZilla:
XGBoost \citep{chen2016xgboost}, linear regression, $k$-nearest neighbors, and uniform random.
For KNN, we set $k=5$ and use the $L_2$ distance from the selected meta-features.

\paragraph{Experimental setup.}
Focusing on performance metric PREC@10, we build a meta-dataset $\mathcal M$
using all rec-sys datasets, algorithms, and meta-features described in Section \ref{sec:analysis}.
We first use the algorithm selection and meta-feature selection procedures described above to select $n=m=100$ parameterized algorithms and meta-features.
For all experiments, we use the best 10 parameterized algorithms selected during this process.
We vary both the number of training meta-datapoints and meta-features; the datapoints and features are randomly selected over 50 random trials.
All RecZilla meta-learners are evaluated using leave-one-dataset-out evaluation: we iteratively select each dataset \emph{family} as the meta-test dataset, and run the full RecZilla pipeline using the remaining datasets as the meta-training data.
Splitting on dataset families rather than datasets ensures that there is no test data leakage.
Then for each dataset $D$ in the test set, we compare the performance metric of the predicted best parameterized algorithm $(a', \bm \omega')$ to the performance metric of the ground-truth best algorithm $y^*$, using the percentage-difference-from-best:
$\%\texttt{Diff} = 100 \times (y^* - \phi(D, a', \bm \omega'))/y^*$.
\%\texttt{Diff} is between 0 and 100, and smaller values indicate better performance.

\begin{table}[t]
\caption{Comparison between RecZilla and two representative algorithm selection approaches from prior work.
To give a fair comparison, the approaches are given the same meta-training datasets.
We compute \%Diff as defined in Section \ref{subsec:experiments}, as well as the Precision@10 for the predicted best algorithm. We report the mean and standard deviation across 50 trials for 19 test sets, for 950 total trials.
The runtime is the average time it takes to output predictions on the meta-test dataset.}
\centering
\begin{tabular}{@{}l|l|l|l@{}}
\toprule
\multicolumn{1}{l}{\textbf{Approach}} & \multicolumn{1}{l}{\textbf{Runtime (sec)}} & \multicolumn{1}{l}{\textbf{\%Diff} ($\downarrow$)} & \multicolumn{1}{l}{\textbf{PREC@10 of best pred.} ($\uparrow$)} \\
\midrule 
\texttt{cunha2018}  \citep{cunha2018metalearning}              & \textbf{0.39}                  & $52.9\pm 23.0$  & $0.00813\pm 0.0113$\\
\texttt{cf4cf-meta} \citep{cunha2018cf4cfmeta}                & 6.68             & $43.5 \pm 21.8$ & $0.00808\pm 0.00773$\\
RecZilla                & 6.69                   & \textbf{33.2}$\pm$ \textbf{22.8} & \textbf{0.00915}$\pm$ \textbf{0.00840} \\
\bottomrule
\end{tabular}
\label{tab:reczilla-comparison}
\end{table}

\paragraph{Results and discussion.}
In Table \ref{tab:reczilla-comparison}, we compare RecZilla with two prior algorithm selection approaches: \texttt{cunha2018} \citep{cunha2018metalearning} and \texttt{cf4c4-meta} \citep{cunha2018cf4cfmeta}, which are a comprehensive depiction of all prior work (see Appendix \ref{subsec:reczilla-comparison-details} for justification, and for details of the experiment). 
Furthermore, note that \texttt{cunha2018} has no open-source code, and \texttt{cf4c4-meta} only has code in R.
We find that RecZilla outperforms the other two approaches in both \%Diff and in terms of the PREC@10 value of the rec-sys algorithm outputted by each meta-learning algorithm.

In Appendix \ref{app:reczilla-ablations},
Figure \ref{fig:performance-reczilla} (left) shows \%\texttt{Diff} vs.\ the size of the meta-training set, and Figure \ref{fig:performance-reczilla} (right) shows the results of an ablation study on the number of selected meta-features $m$.
See Appendix \ref{app:reczilla-ablations} for more details and discussion.

\paragraph{Pre-trained RecZilla models.}
We release pre-trained RecZilla models for PREC@10, NDCG@10, and Item-hit Coverage@10, 
trained with XGBoost on all 18 datasets, with algorithms $n=10$ and meta-features $m=10$.
We also include a RecZilla model that predicts the Pareto-front of PREC@10 and model training time, 
so that users can select their desired trade-off between performance and runtime.
Finally, we include a pipeline so that users can choose a metric from the list of 315 (or any desired combination of the 315 metrics) and train the resulting RecZilla model.

%% file: conclusion.tex
\section{Conclusions, Limitations, and Broader Impact} \label{sec:conclusion}

%

%

In this work, we conducted the first large-scale study of rec-sys approaches: we compared 24 algorithms and 100 sets of hyperparameters across 85 datasets and 315 metrics. We showed that for a given performance metric, the best algorithm and hyperparameters highly depend on the dataset. We also find that various meta-features of the datasets are predictive of algorithmic performance and runtimes.
Motivated by these findings, we created RecZilla, the first metric-independent, hyperparameter-aware algorithm selection approach to recommender systems.
Through empirical evaluation, we show that given a user-defined metric, RecZilla effectively predicts high-performing algorithms and hyperparameters for new, unseen datasets, substantially reducing the need for human involvement.
We not only release our code and pretrained RecZilla models, but we also release the raw experimental results so that users can train new RecZilla models on their own test metrics of interest. This codebase, which includes a unified API for 85 datasets and 24 algorithms, may be of independent interest.

\paragraph{Limitations.}
While our work progresses prior work along several axes, there are still avenues for improvement.
First, the meta-learning problem in RecZilla is low-data. 
Although we added nearly all common rec-sys research datasets into RecZilla, the result is still only 85 meta-datapoints (datasets). 
While we guarded against over-fitting to the training data 
in numerous ways, RecZilla can still be improved by more training data. Therefore, as new recommender system datasets are released in the future, our hope is to add them to our API, so that RecZilla continuously improves over time.
Similarly, our hope is to add the most recent high-performing rec-sys approaches to our work, as well as algorithms released in the future.
This includes adding neural network-based approaches, in addition to the six that we have already included.
%
Another limitation is that RecZilla does not directly predict the performance of hyperparameters 
for algorithms on a given dataset. Although care must be taken to not overfit, modifying RecZilla to
predict the performance of an algorithm together with a set of hyperparameters is an interesting avenue
for future work.
Finally, the magnitude of our evaluation ($84\,850$ models trained) leaves our meta-data susceptible to biases based on experiment success/failures.
While we fixed many common errors such as out-of-memory errors, it was infeasible to give each experiment specific attention. Therefore, RecZilla may have higher uncertainty for the datasets and algorithms that are more likely to fail. An interesting future improvement to RecZilla would be to predict the likelihood that a new dataset will successfully train on a new dataset.

\paragraph{Broader impact.}
Our work is ``meta-research'': there is not one specific application that we target, but our work makes it substantially easier for researchers and practitioners to quickly train recommender system models when given a new dataset.
On the research side, this is a net positive because researchers can much more easily include baselines, comparisons, and run experiments on large numbers of datasets, all of which lead to more principled empirical comparisons.
On the applied side, our day-to-day lives are becoming more and more influenced by recommendations generated from machine learning models, which comes with pros and cons. 
These recommendations connect users with needed items that they would have had to spend time searching for
\citep{jannach2010recommender}.
Although these recommendations may lead to harmful effects such as echo chambers \citep{ge2020understanding, jiang2019degenerate}, 
techniques to identify and mitigate harms are improving \citep{gravino2019towards, morini2021toward}.


%% file: appendix_experiments.tex
\section{Experiment Details}\label{app:experiments}

This appendix outlines the algorithms, datasets, metrics, and hyperparameter selection used in RecZilla, as well as the details of our procedure for generating the meta-dataset.
This codebase is publicly available\footnote{\url{https://github.com/naszilla/reczilla}}, and is written in Python.
The RecZilla codebase builds on another public Github repository\footnote{\url{https://github.com/MaurizioFD/RecSys2019_DeepLearning_Evaluation}}. 

\subsection{Generating Meta-Datasets}\label{app:metadatasets}

We generate the meta-dataset for reczilla using 24 rec-sys algorithms and 85 datasets.
We use a leave-one-out training/validation split for each dataset: for each user, the last interaction is held out for validation, and all remaining interactions are used for training.
For non-neural-network algorithms, each algorithm-dataset pair is given a 10 hour time limit for training and validation and trains/validates with up to 100 random hyperparameter sets (see Appendix~\ref{app:sampling}), on a single a \texttt{n1-highmem-2} instance on Google Cloud (2 vCPUs, 13GB memory). 
Neural-network based methods are trained using the default hyperparameters from their respective papers---meaning we do not sample different hyperparameter sets. 
For each dataset, neural-network methods are run for up to 15 hours on an \texttt{n1-highmem-16} Google Cloud instance (16 vCPUs, 104GB memory), with an NVIDIA Tesla T4 GPU. All neural methods use batch size 64, and are trained for up to 100 epochs; early stoppig occurrs if loss does not improve in 5 epochs.
During validation, we calculate 21 different performance metrics at 15 different cutoffs, for a total of 315 different metrics (see Appendix~\ref{app:metrics}).
Out of all 1\,530 dataset-algorithm combinations tested in our experiments, 1\,404 of them completed the train/validation procedure with at least one hyperparameter set within the 10 hour time limit. 
Most failed experiments failed due to invalid hyperparameter values, and some failed due to memory errors. 

\input{tables/runtime_table}

Algorithm runtime varied substantially across algorithm family and dataset. Table~\ref{tab:alg-runtime} shows runtime statistics over all experiments and all algorithms, for experiments that completed within the 10-hour time limit.

\subsection{Rec-sys Algorithms Implemented in RecZilla} 

Our experiments use 24 rec-sys algorithms.
Algorithms with hyperparameters are associated with a hyperparameter space, as well as a set of ``default'' hyperparameters. 
Table~\ref{tab:algs-params} lists each implemented algorithm, along with its hyperparameter space and default parameters.

In the current implementation, we define several versions of User-KNN and Item-KNN, with one version for each similarity metric. 
This is done for convenience, since different KNN similarity metrics are associated with different hyperparameters.
However, in our experiment results we treat all versions of User-KNN and Item-KNN as the same algorithm.

All algorithms here use the interface from~\cite{DacremaCJ19,dacrema2021troubling} (their codebase is publicly available\footnote{\url{https://github.com/MaurizioFD/RecSys2019_DeepLearning_Evaluation/}}). 
All but two of our 24 algorithms use the implementation from this codebase; two algorithms (CoClustering and SlopeOne) use the implementation of Surprise~\cite{Hug2020}, which is also publicly available.\footnote{\url{http://surpriselib.com/}}

\input{alg_table}

\subsection{RecZilla Datasets}


The RecZilla codebase implements 88 datasets (3 additional datasets were added after our experiments on 85 datasets), derived from 20 dataset families. All datasets are listed in Table \ref{tab:reczilla-datasets}.

{\small
\begin{longtable}[c]{|l|r|r|r|r|}
\caption{Summary of datasets used to train and evaluate RecZilla.
}
\label{tab:reczilla-datasets}\\
\hline
\textbf{Dataset Name}    & \textbf{\# Interactions} & \textbf{\# Items} & \textbf{\# Users} & \textbf{Density}\\ \hline
AmazonAllBeauty          & 232                      & 6,357             & 139               & 2.60E-04                     \\ \hline
AmazonAllElectronics     & 235                      & 7,437             & 124               & 2.50E-04                    \\ \hline
AmazonAlternativeRock    & 328                      & 3,842             & 120               & 7.10E-04                  \\ \hline
AmazonAmazonFashion      & 331                      & 20,800            & 253               & 6.30E-05                    \\ \hline
AmazonAmazonInstantVideo & 75,673                   & 23,965            & 29,756            & 1.10E-04                    \\ \hline
AmazonAppliances         & 2,252                    & 11,402            & 1,581             & 1.20E-04                    \\ \hline
AmazonAppsforAndroid     & 840,985                  & 61,275            & 240,933           & 5.70E-05                    \\ \hline
AmazonAppstoreforAndroid & 19                       & 152               & 16                & 7.80E-03                     \\ \hline
AmazonArtsCraftsSewing   & 97,022                   & 112,334           & 30,712            & 2.80E-05                     \\ \hline
AmazonAutomotive         & 300,532                  & 320,112           & 100,163           & 9.40E-06                     \\ \hline
AmazonBaby               & 236,392                  & 64,426            & 71,826            & 5.10E-05                     \\ \hline
AmazonBabyProducts       & 10,481                   & 9,475             & 5,327             & 2.10E-04                     \\ \hline
AmazonBeauty             & 489,929                  & 249,274           & 146,995           & 1.30E-05                      \\ \hline
AmazonBlues              & 98                       & 896               & 25                & 4.40E-03                      \\ \hline
AmazonBooks              & 11,498,997               & 2,330,066         & 1,686,577         & 2.90E-06                      \\ \hline
AmazonBuyaKindle         & 6,312                    & 1,858             & 2,715             & 1.30E-03                      \\ \hline
AmazonCDsVinyl           & 1,705,140                & 486,360           & 245,080           & 1.40E-05                      \\ \hline
AmazonCellPhonesAccessories	& 588508                & 319,678	        & 245,110	        & 7.51E-06                      \\ \hline
AmazonChristian          & 1,155                    & 7,512             & 428               & 3.60E-04                      \\ \hline
AmazonClassical          & 528                      & 2,301             & 152               & 1.50E-03                      \\ \hline
AmazonClothingShoesJewelry & 1,615,940 	            & 1,136,004 	    & 496,837 	        & 2.86E-06                      \\ \hline
AmazonCollectiblesFineArt	& 1,066 	            & 5,705 	        & 230               &	8.12E-04                     \\ \hline
AmazonComputers          & 51                       & 4,266             & 26                & 4.60E-04                      \\ \hline
AmazonCountry            & 151                      & 1,677             & 47                & 1.90E-03                      \\ \hline
AmazonDanceElectronic    & 686                      & 4,763             & 211               & 6.80E-04                       \\ \hline
AmazonDavis              & 38                       & 58                & 28                & 2.30E-02                       \\ \hline
AmazonDigitalMusic       & 238,151                  & 266,414           & 56,814            & 1.60E-05                       \\ \hline
AmazonElectronics        & 2,302,922                & 476,002           & 651,680           & 7.40E-06                      \\ \hline
AmazonFolk               & 236                      & 2,366             & 38                & 2.60E-03                      \\ \hline
AmazonGiftCards          & 237                      & 345               & 144               & 4.80E-03                      \\ \hline
AmazonGospel             & 105                      & 1,616             & 45                & 1.40E-03                      \\ \hline
AmazonGroceryGourmetFood & 335,994                  & 166,049           & 86,400            & 2.30E-05                      \\ \hline
AmazonHardRockMetal      & 156                      & 1,063             & 41                & 3.60E-03                      \\ \hline
AmazonHealthPersonalCare & 661,968                  & 252,331           & 205,704           & 1.30E-05                      \\ \hline
AmazonHomeImprovement    & 45                       & 3,855             & 32                & 3.60E-04                      \\ \hline
AmazonHomeKitchen        & 1,029,164                & 410,243           & 327,439           & 7.70E-06                      \\ \hline
AmazonIndustrialScientific	& 16,784 	            & 45,383 	        & 7,779 	        & 4.75E-05                       \\ \hline
AmazonInternational      & 608                      & 5,544             & 193               & 5.70E-04                      \\ \hline
AmazonJazz               & 490                      & 2,917             & 109               & 1.50E-03                      \\ \hline
AmazonKindleStore        & 1,387,653                & 430,530           & 213,192           & 1.50E-05                      \\ \hline
AmazonKitchenDining      & 81                       & 3,658             & 63                & 3.50E-04                      \\ \hline
AmazonLatinMusic         & 21                       & 613               & 13                & 2.60E-03                      \\ \hline
AmazonLuxuryBeauty       & 1,564                    & 1,798             & 717               & 1.20E-03                      \\ \hline
AmazonMagazineSubscriptions	& 1,257 	            & 1,422 	        & 560               & 1.58E-03                       \\ \hline
AmazonMiscellaneous      & 416                      & 5,262             & 164               & 4.80E-04                      \\ \hline
AmazonMoviesTV           & 1,894,519                & 200,941           & 319,406           & 3.00E-05                      \\ \hline
AmazonMP3PlayersAccessories	& 19 	                & 1,657 	        & 14                & 8.19E-04                      \\ \hline
AmazonMusicalInstruments & 92,628                   & 83,046            & 29,040            & 3.80E-05                      \\ \hline
AmazonNewAge             & 132                      & 1,276             & 44                & 2.40E-03                       \\ \hline
AmazonOfficeProducts     & 166,878                  & 130,006           & 59,858            & 2.10E-05                      \\ \hline
AmazonOfficeSchoolSupplies	& 41 	                & 3,229 	        & 21 	            & 6.05E-04                      \\ \hline
AmazonPatioLawnGarden    & 134,727                  & 105,984           & 54,196            & 2.30E-05                      \\ \hline
AmazonPetSupplies        & 291,543                  & 103,288           & 93,336            & 3.00E-05                       \\ \hline
AmazonPop                & 435                      & 5,622             & 118               & 6.60E-04                      \\ \hline
AmazonPurchaseCircles    & 17                       & 33                & 11                & 4.70E-02                      \\ \hline
AmazonRapHipHop          & 32                       & 779               & 19                & 2.20E-03                      \\ \hline
AmazonRB                 & 136                      & 2,253             & 69                & 8.70E-04                      \\ \hline
AmazonRock               & 519                      & 4,464             & 97                & 1.20E-03                       \\ \hline
AmazonSoftware           & 29,434                   & 18,187            & 9,097             & 1.80E-04                      \\ \hline
AmazonSportsOutdoors     & 751,440                  & 478,898           & 238,090           & 6.60E-06                      \\ \hline
AmazonToolsHomeImprovement	& 751,440   412,401 	& 260,659 	        & 132,013 	        & 1.20E-05                      \\ \hline
AmazonToysGames          & 549,347                  & 327,698           & 164,590           & 1.00E-05                      \\ \hline
AmazonVideoGames         & 308,086                  & 50,210            & 84,273            & 7.30E-05                      \\ \hline
AmazonWine               & 215                      & 1,228             & 84                & 2.10E-03                      \\ \hline
Anime                    & 7,669,090                & 11,200            & 69,521            & 9.80E-03                             \\ \hline
BookCrossing             & 323,443                  & 340,556           & 22,568            & 4.20E-05                             \\ \hline
CiaoDVD                  & 47,102                   & 16,121            & 4,743             & 6.20E-04                             \\ \hline
Dating                   & 17,088,628               & 168,791           & 135,359           & 7.50E-04                             \\ \hline
Epinions                 & 592,236                  & 139,738           & 28,487            & 1.50E-04                             \\ \hline
FilmTrust                & 32,586                   & 2,071             & 1,336             & 1.20E-02                             \\ \hline
Frappe                   & 17,022                   & 4,082             & 777               & 5.40E-03                             \\ \hline
GoogleLocalReviews       & 4,867,954                & 3,116,785         & 818,824           & 1.90E-06                             \\ \hline
Gowalla                  & 3,735,522                & 1,247,095         & 91,846            & 3.30E-05                             \\ \hline
Jester2                  & 1,640,712                & 140               & 56,333            & 2.10E-01                           \\ \hline
LastFM                   & 89,058                   & 17,632            & 1,883             & 2.70E-03                             \\ \hline
MarketBiasAmazon         & 32,511                   & 9,560             & 20,335            & 1.70E-04                             \\ \hline
MarketBiasModCloth       & 40,633                   & 1,020             & 6,866             & 5.80E-03                             \\ \hline
Movielens100K            & 98,114                   & 1,682             & 943               & 6.20E-02                             \\ \hline
Movielens10M             & 9,833,849                & 10,680            & 69,878            & 1.30E-02                             \\ \hline
Movielens1M              & 986,002                  & 3,882             & 6,039             & 4.20E-02                             \\ \hline
Movielens20M             & 19,723,277               & 27,278            & 138,493           & 5.20E-03                             \\ \hline
MovielensHetrec2011      & 851,372                  & 10,109            & 2,113             & 4.00E-02                             \\ \hline
MovieTweetings           & 808,662                  & 38,018            & 31,917            & 6.70E-04                             \\ \hline
NetflixPrize             & 99,521,398               & 17,770            & 476,694           & 1.20E-02                            \\ \hline
Recipes                  & 819,642                  & 231,637           & 35,464            & 1.00E-04                            \\ \hline
Wikilens                 & 26,316                   & 5,111             & 275               & 1.90E-02                            \\ \hline
YahooMovies	            & 195,947 	                & 11,916 	        & 7,642 	        & 2.15E-03                            \\ \hline
YahooMusic	            & 77,764,403 	            & 98,213 	        & 1,647,758 	    & 4.81E-04                            \\ \hline
\end{longtable}
}

\subsection{Dataset Meta-Features \& Meta-Feature Selection}\label{app:meta-features}

We calculate a total of 383 meta-features for each rec-sys dataset, consisting of a few general meta-features, meta-features describing the distribution of ratings, and performance of landmarkers.

For each dataset, we extract the number of users, number of items, number of ratings, and the ratio of items to users. Furthermore, following the approach outlined in \cite{cunha2016selecting}, we compute the sparsity of the matrix of interactions, and we systematically obtain a series of meta-features based on different distributions that can be obtained by aggregating the ratings in several ways.

\paragraph{Distribution meta-features}

The distribution meta-features are obtained in two steps, as described in \cite{cunha2016selecting}. First, we obtain a distribution. We do this in one of seven ways. We either take all of the ratings at once or we aggreggate ratings for either items or users in one of three different ways: sum, count, or mean. For each of these seven distributions, we then compute ten different descriptive statistics: mean, maximum, minimum, standard deviation, median, mode, Gini index, skewness, kurtosis, and entropy. This results in 70 distribution meta-features.

\paragraph{Landmarkers}

For landmarkers, we evaluate the performance of several baseline algorithms on a subset of the training set. We first select the subsample. Next, we partition this subsample into two sets: a ``sub-training set'' and a ``sub-validation set''. We train each landmarker on the sub-training set and compute performance metrics on the sub-validation set. We compute the 19 performance metrics described in Section \ref{app:metrics}, plus three more algorithm-independent metrics:
\begin{itemize}
    \item \textbf{Items in Evaluation Set:} measures the fraction of items with at least one rating in the evaluation set. This metric is algorithm-independent and only serves to describe the dataset and its split.
    
    \item \textbf{Users in Evaluation Set:} measures the fraction of users with at least one rating in the evaluation set. This metric is algorithm-independent and only serves to describe the dataset and its split.
    \item \textbf{Item Coverage:} fraction of items that are ranked within the top $K$ for at least one user.
\end{itemize}
All 22 metrics are evaluated at cutoffs 1 and 5, to create the meta-features.

The subsampling scheme is designed to satisfy several constraints. We limit the number of users to 100 and the number of items to 250. We also need to ensure there are at least 2 items rated per user so that the subsequent data split (holding out one item rating per user) does not result in cold users. Furthermore, we ensure the number of items selected is at least 6 so that we can evaluate the performance metrics with the cutoff of 5.

We start the subsampling process by filtering by the users that have at least 2 ratings. We next filter by the items all those users rated. If this results in less than 6 items, we add a random sample of the remaining (cold) items back to the set to have 6 items in total. Next, if the number of users is larger than 100, we take a random subsample of 100 of them and filter by those users. As before, we filter by the items those users rated, and ensure this results in at least 6 items by adding random items back if needed. If the number of items is still greater than 250, we must build an item subsample that results in at least two ratings per user. For each user, we randomly choose two of the items the user rated, and we take the union of these item choices. If this results in less than 250 items, we take a random sample of the remaining items to make the total number of items equal to 250.

Once this subsample is built, we split the subsample into the sub-training and sub-validation set by leaving 1 random item out for each user. We are then ready to run our landmarkers on those sets.

Our landmarkers consist of TopPop, ItemKNN, UserKNN, and PureSVD. For ItemKNN and UserKNN, we use cosine similarity and set the number of neighbors $k$ to 1 and 5. For PureSVD, we use 1 and 5 for the number of latent factors. This results in a total of 7 landmarkers.

Running all 7 landmarkers and computing all 22 metrics at the 2 different cutoffs results in a total of 308 landmarker meta-features.

\subsection{Hyperparameter Sampling}\label{app:sampling}

For each algorithm with hyperparameters, we test up to 100 parameter sets, limited by the 10-hour time limit used in our experiments. 
The first evaluated hyperparameter set is the default hyperparameters
\footnote{See \url{https://github.com/naszilla/reczilla}.}
The remaining 99 hyperparameters are sampled using the ranges specified in Table \ref{tab:algs-params}, using Sobol sampling.

\subsection{Evaluation Metrics}\label{app:metrics}

Each of the evaluation metrics that we use during validation measures the quality of ranking based on the item ratings. For each user, we generate predicted ratings for all items and rank the items according to the predicted rating (in descending order). We trim these ranked lists at a given cutoff, which we denote by $K$. We then compute different metrics using these user-wise top $K$ items. Some metrics also consider the set of relevant items within these top $K$, defined as those for which the user rated the item in the evaluation set.

We use 21 different metrics, and we compute them using cutoffs in $\{1, 2, 3, 4, 5, 6, 7, 8, 9, 10, 15, 20, 30, 40, 50\}$. This results in 315 different metric/cutoff combinations total. Only a small subset of these metrics are used in our analysis; however, any user-chosen metric (or combination of metrics) can be used to define a performance function for RecZilla.

The metrics are calculated using the implementation in the public repository\footnote{\url{https://github.com/MaurizioFD/RecSys2019_DeepLearning_Evaluation}} that our codebase is built on.
Below is a list of metrics calculated during model evaluation:
\begin{itemize}
    \item \textbf{Average Popularity:} measures the popularity of the recommended items. The popularity for each item is the frequency with which it was rated in the training set. These popularities are normalized by the largest popularity. Next, for each user, we compute the normalized popularity of each of the items within its top $K$ and take their mean. Finally, we average across all users.
    \item \textbf{Average Reciprocal Hit-Rank (ARHR):} similar to MRR, except the reciprocal ranks for all relevant items (not just the first) are summed together.
    \item \textbf{Diversity (Gini):} computes the Gini diversity index \footnote{\url{https://www.statsdirect.com/help/default.htm\#nonparametric_methods/gini.htm}} of the global distribution of items ranked within the top $K$ across all users. Higher values indicate higher diversity.
    \item \textbf{Diversity (Herfindahl):} computes the Herfindahl index \cite{adomavicius2011improving} of the global distribution of items ranked within the top $K$ across all users. Higher values indicate higher diversity.
    \item \textbf{Diversity (Shannon):} computes the Shannon entropy of the global distribution of items ranked within the top $K$ across all users.
    \item \textbf{F1 Score:} the harmonic mean between precision and recall.
    \item \textbf{Hit Rate:} fraction of users for which at least one relevant item is present within the top $K$.
    \item \textbf{Item Coverage (Hit):} fraction of relevant items that are ranked within the top $K$ for at least one user.
    \item \textbf{Mean Average Precision (mAP):} the mean of the average precision across all users. The average precision for a user is computed as follows: for any position $i\leq K$ occupied by a relevant item, we compute the precision at $i$. We sum all of these precision values and divide the total by $K$.
    \item \textbf{Mean Average Precision - Min Den:} similar to mAP, but using a modified version of the average precision. If the number of test items for the user is smaller than $K$, we divide the sum (in the last step of the average precision computation) by this number instead of by $K$.
    \item \textbf{Mean Inter-List Diversity:} measures how different the top $K$ lists are for all users, as originally proposed in \cite{zhou2010solving}. For each pair of users, we compute the fraction of items in their top $K$ items that are not present in both lists. Taking the average across all users yields the mean inter-list diversity. The codebase implements a more efficient but equivalent way of computing this metric.
    \item \textbf{Mean Reciprocal Rank (MRR):} the mean of the reciprocal rank across all users. The reciprocal rank for a user is the reciprocal of the rank of the most highly-ranked relevant item or 0 if there is none.
    \item \textbf{Normalized Discounted Cumulative Gain (NDCG):} first, the discounted cumulative gain (DCG) of the top $K$ ranking is computed by adding, for all relevant items in the top $K$, a gain discounted logarithmically in terms of the rank. The NDCG is obtained by dividing the DCG of the ranking by that of an ideal ranking (a ranking ordered by relevance).
    \item \textbf{Novelty:} a metric that rewards recommending items that were not popular in the training set \cite{zhou2010solving}. For each item, we compute the fraction of ratings in the training set that correspond to the item. The novelty contributed by the item is computed by taking the negative logarithm of that fraction and dividing by the total number of items, so that items that were seldom seen in the training set result in high contributions. Now, for any user, we compute the novelty as the sum of these contributions for the top $K$ items. Finally, we average the metric across all users.
    \item \textbf{Precision:} the fraction of items in the top $K$ that are relevant, computed across all users.
    \item \textbf{Precision Recall Min:} similar to precision, but if there are less test items than $K$, the fraction is computed with respect to the number of test items.
    \item \textbf{Recall:} the fraction of relevant items that were placed within the top $K$, computed across all users.
    \item \textbf{User Coverage:} fraction of users both present in the evaluation set and for which the model is able to generate recommendations. In practice, all the algorithms used were able to generate recommendations for all users, since our splitting procedures did not result in cold users, so this metric was always 1 for our datasets and algorithms.
    \item \textbf{User Coverage (Hit):} fraction of users both present in the evaluation set and for which the model is able to generate at least one relevant recommendation within the top $K$ items. It is equal to the product of the hit rate and the user coverage. Because the latter was always equal to 1, the user coverage (hit) was the same as the hit rate in our experiments.
    
    
\end{itemize}

\subsection{Additional details and experiments from Section \ref{subsec:predictability}}

Recall that in Table \ref{tab:top-corrs}, we showed the meta-features that are most highly correlated with the performance (PREC@10) of each algorithm, using their default parameters. In Table \ref{tab:top-corrs-time}, we run the same analysis using ``training time'' instead of PREC@10 as the metric.
We see that for some algorithms, the runtime is very highly correlated with certain meta-features such as ``number of interactions''.

\begin{table}[t]
\caption{Highest absolute correlations between algorithm running time (default hyperparameters) and meta-features.}
\centering
\begin{tabular}{@{}l|l|l@{}}
\toprule
\multicolumn{1}{l}{\textbf{Abs. Correlation}} & \multicolumn{1}{l}{\textbf{Algorithm Family}} & \multicolumn{1}{l}{\textbf{Meta-feature}} \\
\midrule 
0.999                & MF-FunkSVD                  & Number of interactions   \\
0.997                & MF-BPR             & Number of users \\
0.993                & GlobalEffects                   & Number of interactions   \\
0.990                & TopPop                   & Number of interactions  \\
0.986                & ItemKNN                   & Kurtosis of item rating sum distribution  \\
\bottomrule
\end{tabular}
\label{tab:top-corrs-time}
\end{table}

Next, we compute a simple measure of \emph{dataset hardness}, which we compute as, given a performance metric, the negative of the maximum value achieved for that dataset across all algorithms. For example, if all 20 algorithms do not perform well on the MovieTweetings dataset, then we can expect that the MovieTweetings dataset is ``hard''.
In Table \ref{tab:dataset-hardness}, we show the meta-features that are most highly correlated with the \emph{hardness} of each algorithm, where hardness is calculated as -PREC@10. 
We find that the entropy of the rating matrix is most correlated with dataset hardness.

\begin{table}[t]
\caption{Highest absolute correlations between dataset hardness (negative of maximum PREC@10 achieved over all algorithms) and meta-features.}
\centering
\begin{tabular}{@{}l|l|l@{}}
\toprule
\multicolumn{1}{l}{\textbf{Abs. Correlation}} & \multicolumn{1}{l}{\textbf{Pos. or Neg. Corr.}} & \multicolumn{1}{l}{\textbf{Meta-feature}} \\
\midrule 
0.752                & Negative                  & Entropy of ratings  \\
0.668                & Negative             & Mode of user rating count distribution \\
0.655                & Negative                   & Landmarker, UserKNN ($k=5$), Item Coverage @ 1 \\
0.652                & Negative                   & Landmarker, TopPop, Recall @ 5  \\
0.652                & Negative                   & Landmarker, TopPop, Hit Rate @ 5  \\
\bottomrule
\end{tabular}
\label{tab:dataset-hardness}
\end{table}


\paragraph{Additional details from Section \ref{subsec:predictability}.}
In this section, we give more details of the experimental setup of the experiment in Figure \ref{fig:meta-learner} from Section \ref{subsec:predictability}.
We compare three different meta-learner functions: XGBoost, linear regression, and KNN with $k$=5 and $L_2$ distance.
In order to give a fair comparison, we use a fixed set of 10 high-performing rec-sys algorithms, and also a fixed set of 10 meta-features that have high correlation with the performance metric PREC@10.
The meta-learner models are trained to predict PREC@10 of all 10 algorithms.

We use leave-one-out evaluation for each meta-learner: one dataset family is held out for testing, while $x$ other dataset families are used for training, for $x$ from 2 to 20.
For each meta-learner, we compute the mean absolute error (MAE): the absolute differences between the prediction and ground truth for each algorithm are averaged. These are then averaged over 200 trials of randomly-selected training dataset families.

%% file: tables/runtime_table.tex
\begin{table}
    \caption{Min, mean, and max runtime for each algorithm, over all experiments, for both training and evaluation. The rightmost column shows the number of experiments collected for each algorithm. These runtime statistics only include experiments that completed within the 10 hour time limit, so they are skewed to be small, and should be interpreted as general trends. GlobalEffects, SlopeOne, Random, and TopPop do not have hyperparameters and therefore completed a maximum of 85 experiments. Furthermore, we tested multiple distance metrics for Item-KNN and User-KNN, resulting in more experiments.
    \label{tab:alg-runtime}}
\begin{tabular}{lrrrrrrr}
\toprule
{} & \multicolumn{3}{l}{Training time (seconds)} & \multicolumn{3}{l}{Evaluation time (seconds)} & Num. experiments\\
{} &           min &   mean &      max &          min &   mean &      max &       size \\
Alg. family                         &               &        &          &              &        &          &            \\
\midrule
CoClustering                       &          0.02 & 167.45 & 25766.23 &         0.05 &  51.09 & 11393.36 &       5106 \\
EASE-R                 &          $<$0.01 &  13.85 &   454.14 &         0.05 &  14.99 &   341.97 &       4376 \\
GlobalEffects                      &          $<$0.01 &   0.43 &    12.73 &         0.05 & 639.58 &  8191.36 &         85 \\
iALS                    &          0.69 & 369.30 & 24283.54 &         0.05 &  10.26 &  3558.97 &       3502 \\
Item-KNN                          &          $<$0.01 & 100.28 & 13617.51 &         0.05 & 159.03 &  8192.90 &      12847 \\
MF-AsySVD  &          0.03 & 207.76 & 28463.53 &         0.05 &  40.37 & 13293.74 &       4254 \\
MF-BPR     &          0.02 &  82.19 &  9635.26 &         0.06 &  69.99 & 12117.41 &       5659 \\
MF-FunkSVD &          0.02 & 172.61 & 15650.11 &         0.05 &  50.49 & 12793.54 &       4938 \\
NMF                     &          0.01 & 221.50 & 20369.45 &         0.06 &  87.42 & 11471.30 &       2957 \\
P3alpha                &          $<$0.01 &  78.17 &  6264.47 &         0.05 &  62.40 &  6563.46 &       5816 \\
SVD                &          $<$0.01 &   3.58 &   353.12 &         0.05 &  99.06 & 10393.31 &       6132 \\
RP3beta                &          $<$0.01 &  80.45 &  7043.17 &         0.05 &  61.24 &  7067.75 &       5900 \\
Random                             &          $<$0.01 &   0.09 &     2.35 &         0.06 & 937.67 & 16529.64 &         85 \\
SLIME-lasticNet          &          0.06 & 142.95 & 25731.74 &         0.05 &  11.63 &  1816.69 &       4706 \\
SLIM-BPR                    &          0.03 &  82.57 & 31063.87 &         0.05 &  21.89 &  1962.25 &       5176 \\
SlopeOne                           &          0.05 &  17.73 &   470.27 &         0.07 &  45.77 &   803.25 &         48 \\
TopPop                             &          $<$0.01 &   0.13 &     3.66 &         0.05 & 626.50 &  7501.84 &         85 \\
User-KNN                          &          $<$0.01 &  73.28 & 30975.03 &         0.05 & 158.07 &  6327.46 &      13097 \\
\bottomrule
\end{tabular}
\end{table}

%% file: alg_table.tex
{\small
\begin{longtable}[c]{l|m{3cm}|l}
    \caption{Description of all algorithms implemented in RecZilla.}
    \label{tab:algs-params}\\
    \toprule
        Algorithm Name & Reference/ Description& Hyperparameter Space  \\ \midrule
    CoClustering  & Clusters users and items. Uses their average ratings to predict new ratings. \cite{george2005scalable} \cite{Hug2020} & $\begin{array}{S@{:\,}L} 
num-control-users&  Int(1, 1000)\\
num-control-items&  Int(1, 1000)\end{array}$\\ \midrule
    EASE-R  & Linear model designed for sparse data. Simplified version of an autoencoder~\cite{steck2019embarrassingly}. & 
    $\begin{array}{S@{:\,}L} l2-norm& [1, 1e7]
                \end{array}$
                \\ \midrule
    GlobalEffects  & Rating predictions are based on a global score for each item and each user. & - \\ \midrule
    iALS  & Matrix factorization method. Leverages alternating least squares for optimization and uses regularization~\cite{hu2008collaborative}. & 
    $\begin{array}{S@{:\,}L} num-factors& Int(1, 200) \\
                confidence-scaling& \{lin., log.\} \\
                alpha& [1e-3, 50] \\
                epsilon& [1e-3, 10] \\
                reg& [1e-5, 1e-2]\\
                 \end{array}$ \\ \midrule
     ItemKNN-Asymmetric  &  k-nearest neighbors, item-based.~\cite{itemknn,dacrema2021troubling} Similarity between items is calculated using the asymmetric cosine similarity~\cite{asymcos}. & $\begin{array}{S@{:\,}L} top-K & Int(5, 1000) \\
    shrink& Int(0, 1000) \\ 
    alpha& [0, 2] \end{array}$   \\ \midrule
    ItemKNN-Cosine  & k-nearest neighbors, item-based.~\cite{itemknn,dacrema2021troubling} Similarity between items is calculated using the cosine similarity. & $\begin{array}{S@{:\,}L} top-K & Int(5, 1000) \\ 
    shrink& Int(0, 1000) \\
    normalize& Bool \\
    feature-weighting& \{none, BM25, TF-IDF\} \end{array}$    \\ \midrule
    ItemKNN-Dice  & k-nearest neighbors, item-based.~\cite{itemknn,dacrema2021troubling} Similarity between items is calculated using the S\o{}rensen-Dice coefficient.~\cite{dice} & $\begin{array}{S@{:\,}L} top-K & Int(5, 1000) \\
    shrink& Int(0, 1000) \\
    normalize& Bool \end{array}$     \\ \midrule
    ItemKNN-Euclidean  & k-nearest neighbors, item-based.~\cite{itemknn,dacrema2021troubling} Similarity between items is calculated using the euclidean distance (l2 distance). & $\begin{array}{S@{:\,}L} top-K & Int(5, 1000) \\ 
shrink& Int(0, 1000) \\
normalize& Bool \\ 
normalize-avg-row& Bool  \\ 
similarity-from-distance& \{lin., log., exp.\} \end{array}$   \\ \midrule
    ItemKNN-Jaccard  & k-nearest neighbors, item-based.~\cite{itemknn,dacrema2021troubling} Similarity between items is calculated using the Jaccard index.~\cite{murphyjaccard} & Same as ItemKNN-Dice    \\ \midrule
    ItemKNN-Tversky  &
    k-nearest neighbors, item-based.~\cite{itemknn,dacrema2021troubling} Similarity between items is calculated using the Tversky index.~\cite{tversky1977features}& $\begin{array}{S@{:\,}L} top-K & Int(5, 1000) \\ 
    shrink& Int(0, 1000) \\
    alpha& [0, 2] \\
    beta& [0, 2] \end{array}$    \\ \midrule
    MF-AsySVD  & Matrix factorization model that replaces user factors with the factors of items rated by that user. Items have multiple corresponding factors. \cite{asysvd}  & 
    $\begin{array}{S@{:\,}L} 
                sgd-mode& \{sgd, adagrad, adam\} \\
                use-bias& Bool \\
                num-factors& Int(1, 200) \\
                item-reg& [1e-5, 1e-2] \\
                user-reg& [1e-5, 1e-2] \\
                learning-rate& [1e-4, 1e-1]\\
                negative-interactions-quota& [0, 0.5]
                 \end{array}$    \\ \midrule
    MF-BPR  & Uses the Bayesian Personalized Ranking loss to learn a matrix factorization model~\cite{rendle2012bpr}. & 
    $\begin{array}{S@{:\,}L} sgd-mode& \{sgd, adagrad, adam\} \\
                num-factors& Int(1, 200) \\
                batch-size& \{1, 2, 4, 8, 16, 32, 64, 128, 256, 512, 1024\} \\
                positive-reg& [1e-5, 1e-2] \\
                negative-reg& [1e-5, 1e-2] \\
                learning-rate& [1e-4, 1e-1]\\
                 \end{array}$  \\ \midrule
    MF-FunkSVD & A modified version of the matrix factorization algorithm proposed in a blog post\footnote{\url{https://sifter.org/~simon/journal/20061211.html}}~\cite{dacrema2021troubling}.  & $\begin{array}{S@{:\,}L} sgd-mode& \{sgd, adagrad, adam\} \\
                use-bias& Bool \\
                batch-size& \{1, 2, 4, 8, 16, 32, 64, 128, 256, 512, 1024\} \\
                num-factors& Int(1, 200) \\
                item-reg& [1e-5, 1e-2] \\
                user-reg& [1e-5, 1e-2] \\
                learning-rate& [1e-4 1e-1] \\
                negative-interactions-quota& [0, 0.5] \end{array}$  \\ \midrule
    NMF  & Non-negative matrix factorization~\cite{cremonesi2010performance}. & 
        $\begin{array}{S@{:\,}L} num-factors& Int(1, 350) \\
                    solver& \{coordinate-descent, multiplicative-update.\} \\
                    init-type& \{random, nndsvda\} \\
                    beta-loss& \{frobenius, kullback-leibler\}
                     \end{array}$ \\ \midrule
    P3alpha & Computes the relevance between users and items based on random walks in a graph containing both users and items~\cite{cooper2014random}.  &  $\begin{array}{S@{:\,}L} top-K & Int(5, 1000) \\
    alpha& [0, 2] \\
    normalize-similarity& Bool \end{array}$    \\  \midrule
    PureSVD  & Matrix factorization method based on SVD. &  $\begin{array}{S@{:\,}L} num-factors& Int(1, 200) \end{array}$\\ \midrule
    Random  & Predicts random ratings. & - \\ \midrule
    RP3beta & Similar to P3alpha, but uses a reweighing scheme to compensate for item popularity~ \cite{paudel2016updatable}.  & $\begin{array}{S@{:\,}L} top-K & Int(5, 1000) \\
    alpha& [0, 2] \\
    beta& [0, 2] \\
    normalize-similarity& Bool \end{array}$  \\ \midrule
    SLIM-BPR  & Uses a Sparse Linear Method (SLIM) optimized for Bayesian Personalized Ranking (BPR) loss. \cite{slim-bpr,dacrema2021troubling} &
    $\begin{array}{S@{:\,}L} top-K& Int(5, 1000) \\
                symmetric& Bool \\
                sgd-mode& \{sgd, adagrad, adam\} \\
                 lambda-i& [1e-5, 1e-2] \\
                  lambda-j& [1e-5, 1e-2] \\
                learning-rate& [1e-4, 1e-1]\\
                \end{array}$
    \\ \midrule
    SLIMElasticNet  & Sparse Linear Method (SLIM) \cite{levy2013efficient,dacrema2021troubling}  & 
    $\begin{array}{S@{:\,}L} top-K& Int(5, 1000) \\
                symmetric& Bool \\
               l1-ratio& [1e-5, 1] \\
               alpha& [1e-3, 1e-2] 
                \end{array}$
    \\  \midrule
    SlopeOne  & Uses linear functions to predict ratings for an item based on those from other items. \cite{lemire2005slope} \cite{Hug2020} &  - \\ \midrule
    TopPop  & Recommends items based on global popularity regardless of user. & - \\  \midrule
    UserKNN-Asymmetric  & k-nearest neighbors, item-based, using the asymmetric cosine similarity.~\cite{userknn,dacrema2021troubling} & Same as ItemKNN-Asymmetric   \\ \midrule
    UserKNN-Cosine  &  k-nearest neighbors, user-based, using the cosine similarity.~\cite{userknn,dacrema2021troubling}& Same as ItemKNN-Cosine   \\  \midrule
    UserKNN-Dice  & k-nearest neighbors, user-based, using the S\o{}rensen-Dice coefficient.~\cite{userknn,dacrema2021troubling} & Same as ItemKNN-Dice    \\ \midrule
    UserKNN-Euclidean  & k-nearest neighbors, user-based, using the euclidean distance.~\cite{userknn,dacrema2021troubling} & Same as ItemKNN-Euclidean    \\  \midrule
    UserKNN-Jaccard  & k-nearest neighbors, user-based, using the Jaccard index.~\cite{userknn,dacrema2021troubling} & Same as ItemKNN-Jaccard    \\ \midrule
    UserKNN-Tversky  & k-nearest neighbors, user-based, using the Tversky index.~\cite{userknn,dacrema2021troubling}  & Same as ItemKNN-Tversky    \\ \bottomrule
\end{longtable}
}

%% file: appendix_reczilla_pipeline.tex
\section{RecZilla Meta-Learning Pipeline}
In this section, we give more details of the RecZilla pipeline, and we give an additional experiment in which we compare RecZilla to other existing rec-sys meta-learning approaches.

The RecZilla pipeline consists of the following components: initial algorithm selection, meta-feature selection, and finally the meta-learner for algorithm selection. The purpose of the first two components is to reduce the dimensionality of the dataset and reduce the risk of overfitting for the classifier. We describe each of these components in the following sections. In all that follows, we assume that the user provides (a) a performance metric function $\phi$, (b) the number of parameterized algorithms to be considered by the meta-learner $n$, and (c) the number of dataset meta-features to be considered by the meta-learner $m$.
In addition, we assume access to a meta-dataset $\mathcal M$, such as the one described (and already pre-computed) in this paper.

\subsection{Initial algorithm selection}

We first select $n$ parameterized algorithms which have high \emph{coverage} over all datasets in meta-dataset $\mathcal M$ (see Section~\ref{sec:reczilla-pipeline} of the paper).
Since data is relatively scarce in this meta-learning task, we select a subset of $n$ algorithms to reduce the dimensionality of the meta-learner prediction target.

\subsection{Meta-feature selection}

Since our meta-dataset $\mathcal M$ includes hundreds of features, we restrict our meta-learner to $m$ features to avoid over-fitting.
This is the same approach taken by prior work \citep{cunha2018metalearning}.
It is computationally infeasible to find the set of $m=10$ best meta-features out of a set of 383, since we would need to check ${383 \choose 10}\approx 10^{19}$ combinations of meta-features.
Instead, we iteratively grow a set of $m$ meta-features which are highly correlated with the user-specified performance metric, without selecting redundant features, by
using a ``greedy'' approach (similar in spirit to prior work \citep{cunha2018metalearning}).
Here we require that the (a) performance metric function $\phi$ is chosen ahead of time, and (b) a set of $n$ parameterized algorithms have been selected. 
We introduce some additional notation for this section to describe our meta-feature selection process.
Let $\bm y^i \in \mathbb R^{|\mathcal D|}$ be the vector of performance metric $\phi$ for parameterized algorithm $i\in 1, \dots, n$, for all datasets in $\mathcal D$.
Let $\bm d^j\in \mathbb R^{|\mathcal D|}$ be the vector of meta-feature $j$ for all datasets in $\mathcal D$, and let $J$ be the total number of meta-features.
Let $F$ denote a set of feature indices that corresponds to selected features. 

For each (algorithm, meta-feature pair), we first compute the absolute value of the Pearson correlation between the meta-feature and the performance of each parameterized algorithm $i$, across all datasets: $c_{ij} \gets |\text{corr}(\bm y^i, \bm d^j)|$ for all $i\in \{ 1, \dots, n\}=[n]$ and $j\in \{1, \dots, J\}=[J]$. When computing this correlation, each sample is weighed inverse-proportionally to the size of the dataset family it corresponds to---to prevent large dataset families (such as Amazon) from dominating the correlation computation.

We select the first feature by finding the largest absolute correlation coefficient between any of the meta-features and parameterized algorithms, and we choose the meta-feature corresponding to it: 
$$j_0 \gets \mathop{\arg\max}\limits_{j\in [J]}\left(\max_{i\in [n]} c_{ij}\right).$$
All remaining $(m-1)$ meta-features are selected such that we maximize the \emph{improvement} in the absolute correlation between the selected features and the selected algorithms' performance.
This way, we avoid selecting highly correlated features.
Algorithm~\ref{alg:meta-feature-selection} gives a pseudocode description of this feature-selection process. 

\begin{algorithm}
\caption{RecZilla Meta-feature selection}\label{alg:meta-feature-selection}
\begin{algorithmic}
\Require$\bm d^j\in \mathbb R^{|\mathcal D|},\, \forall j\in [J]$ \Comment{$J$ vectors of meta-features for each dataset}
\Require $\bm y^i\in \mathbb R^{|\mathcal D|},\, \forall i\in [n]$ \Comment{$n$ vectors of performance metrics for each algorithm}
\Require $m>0$ \Comment{number of features to select}
\State $F\gets \{\}$ \Comment{indices of selected features}
\State $x_i=0,\, \forall i\in [n]$ \Comment{max abs. correlation between any selected meta-feature and $\bm y^i$}
\State $c_{ij}\gets |\text{corr}(\bm y^i, \bm d^j)|,\, \forall i\in [n], j\in [J]$
\While{$|F|<m$}
    \State $j' \gets \mathop{\arg\max}\limits_{j\in [J]}\left[\max\limits_{i\in [n]} (c_{ij} - x_i) \right]$
    \State $F \gets F \cup \{j'\}$
    \State $x_i \gets \max\{x_i, c_{ij'}\},\,\forall i\in [n]$ \Comment{update the max. abs. correlation for each alg.}
\EndWhile

\State \Return $F$
\end{algorithmic}
\end{algorithm}

\subsection{Metalearner for algorithm selection}

The goal of the metalearner is to predict the performance of all $n$ selected parameterized algorithms on a new dataset. The input to this meta-learner is the set of $m$ meta-features selected in the previous selection, and the output is an $n$-dimensional vector of performance metrics for all selected algorithms.
We treat this as a multi output regression problem, and our experiments test three different models: a RegressiorChain with XGBoost as the base model, KNN with $k=5$ and $L_2$ distance, and multinomial linear regression. For training these meta-learner models, we used the squared error cost function. 

To train the meta-learner we construct a final meta-dataset consisting of one tuple $(\bm d, \bm y)$ for each dataset represented in $\mathcal M^{train}$, where $\bm d$ is a vector of $m$ meta-features for the dataset, and $\bm y$ is a vector of the performance of $n$ parameterized rec-sys  algorithms in the set of selected parameterized algorithms $\mathcal S'$.

\subsection{Additional details on comparisons to other algorithm selection methods for rec-sys} \label{subsec:reczilla-comparison-details}
As described in section of Section \ref{sec:introduction}, there are a few existing works on algorithm selection for recommender systems. 
In this section, we describe the two most relevant and recent works in more detail, and we give more details on the experiment in Section \ref{sec:reczilla} that empirically compared RecZilla with existing works (Table \ref{tab:reczilla-comparison}).

Multiple approaches for algorithm selection for recommender systems were developed between 2011 and 2018 \citep{huang2011does, ekstrand2012recommenders, adomavicius2012impact, griffith2012investigations, cunha2016selecting},
but in 2018, Cunha et al.\ \citep{cunha2018metalearning} gave a thorough empirical analysis of all recommender systems studied up to that point.
They showed that the best approach was to use meta-feature selection over 74 distribution-based meta-features, and to use polynomial SVM as the meta-model.
Their study used 38 total datasets.
We refer to this algorithm as \texttt{cunha2018}.

Two more algorithm selection for rec-sys approaches were released after Cunha et al.\ \citep{cunha2018metalearning}.
CF4CF \citep{cunha2018cf4cf}, which used subsampling landmarkers, and CF4CF-META, which combined CF4CF with the earlier meta-learning approaches described in the previous paragraph.
Specifically, CF4CF-META uses landmarkers as well as the 74 distribution-based meta-features. It also uses KNN as the meta-model. 
Their study uses 38 total datasets.
We refer to this approach as \texttt{cf4cf-meta}.

Given that \texttt{cunha2018} subsumes all work prior to 2018, and \texttt{cf4c4-meta} subsumes all subsequent work, we compare RecZilla against these two algorithms as a comprehensive depiction of all prior work. Note that 
\texttt{cunha2018} has no open-source code, and \texttt{cf4c4-meta} only has code in R.
Furthermore, in order to give a more fair empirical study, we implement both approaches directly within our codebase. Additionally, to give a fair comparison, each model uses the same meta-training datasets, algorithm selection procedure, and base algorithms.
Since a main novelty of RecZilla is predicting hyperparameters as well as algorithms, the other two approaches are only given the algorithms with the default hyperparameters.

We run an experiment with \texttt{cunha2018}, \texttt{cf4c4-meta}, and RecZilla, in the same setting as Section \ref{subsec:experiments} (i.e., the setting of Figure \ref{fig:performance-reczilla}), where we run leave-one-dataset-out evaluation and average the results over all test datasets. The algorithms are given all 18 dataset families not in the test set, to use as training data.
We run 50 trials for all 19 possible test sets in the leave-one-dataset-out evaluation, for a total of 950 trials.
See Table \ref{tab:reczilla-comparison}.
RecZilla outperforms the other two approaches in both \%Diff and in terms of the PREC@10 value of the rec-sys algorithm outputted by each meta-learning algorithm.

%% file: appendix_results.tex
\section{Additional Results and Discussion}\label{app:results}

\subsection{Detailed algorithmic results}

Tables \ref{tab:all-alg-ranks}, \ref{tab:all-alg-ranks-b}, and \ref{tab:all-alg-ranks-c} show the min (best), max (worst) and mean rank for each algorithm, over all 85 datasets, over a subset of metrics.
Most algorithms are ranked in the top-3 for each metric, on at least one dataset; furthermore, most algorithms are ranked \emph{poorly} on at least one dataset.
However, average performance varies widely across algorithms; for example, Item-KNN is has an average rank at most 3 for most metrics, while TopPop, GLobalEffects, SlopeOne, have average ranks closer to 20.

\input{tables/table_2a_appendix}
\input{tables/table_2b_appendix}
\input{tables/table_2c_appendix}

\subsection{A guide to practitioners}
In this section, we lay out the key takeaways and insights for practical recommender system design, so that practitioners can get the largest value from our work and use our work most effectively.

\paragraph{Insights from Section \ref{sec:analysis}.}
In Section \ref{sec:analysis}, we presented a large-scale empirical study of rec-sys algorithms, studying the generalizability and predictability of rec-sys algorithms.
The key takeaway from this section for practitioners is that rec-sys algorithms are \emph{not} generalizable (see Table \ref{tab:overall-ranks} and Figure \ref{fig:squiggly}).
Therefore, practitioners who have used an algorithm successfully on one rec-sys dataset will likely need to try other diverse types of algorithms when starting on a brand new dataset, even if using a neural network-based algorithm.

Practitioners may also use our analysis for intuition when working with RecZilla models, in a number of different ways. 
The particular use cases and goals of a practitioner may be very specific, but we give a few examples on how to leverage insights from Section \ref{sec:analysis} below.
\begin{itemize}
    \item Table \ref{tab:dataset-hardness} gives properties of a dataset that are predictive of ``dataset hardness.''
    Of course, this is based on correlation alone, but it suggests that a dataset is ``harder'' than average if the \texttt{entropy of ratings} is greater than 2.07 (this value can be computed directly inside our codebase
    \footnote{See \url{https://github.com/naszilla/reczilla/blob/main/RecSys2019_DeepLearning_Evaluation/Metafeatures/README.md} and \url{https://github.com/naszilla/reczilla/blob/main/RecSys2019_DeepLearning_Evaluation/Metafeatures/DistributionFeatures.py}
    }).
    If less than 2.07, the practitioner can continue as usual, but if greater than 2.07, it may be desirable to check if the dataset is noisy and can be cleaned, or if more data can be obtained.
    Or, it could be a sign that the practitioner should expect to spend more time than usual tuning the recommender system model (after running RecZilla, of course).
    \item If the practitioner is concerned with the training time and latency of their recommender system model, then they can estimate the runtimes in advance.
    For example, according to Table \ref{tab:alg-runtime}, \texttt{number of interactions} has a very high correlation with the training time of \texttt{MF-FunkSVD}. 
    By running a simple regression, we have the following formula:
    \begin{equation*}
        \texttt{runtime of MF-FunkSVD} = 0.00439 \cdot (\texttt{number of interactions}) + 0.304.
    \end{equation*}
    Similar formulas can be computed for other algorithms, by using our codebase$^{11}$. 
    \item If the practitioner is interested in using a specific algorithm on a variety of datasets, then they can gain insights by looking up dataset meta-features that specifically have high correlation with the performance metric and runtime of that algorithm.
    For example, a practitioner might prefer to use the CoClustering algorithm for the interpretable clusters that it gives.
    By looking at Table \ref{tab:top-corrs}, and by computing additional correlations using our codebase$^{11}$, the practitioner can find the insight that \texttt{Median of item rating count distribution} and \texttt{Median of item rating sum distribution} are both predictive of the performance of CoClustering, which may be useful to know as a sanity check when tuning and deploying CoClustering on different datasets.
\end{itemize}

\paragraph{A guide for using RecZilla.}
The goal of RecZilla is to allow practitioners to very quickly train a model that performs well, when faced with a new dataset. The practitioner may then choose to further tune the model to reach even stronger performance.

The first step for the practitioner is to define their objective. If the objective is PREC@10, NDCG@10, or Item-hit coverage@10, then the practitioner can use one of the pre-trained models included in our repository.
First, follow the installation instructions in the README.
\footnote{\url{https://github.com/naszilla/reczilla}}
Next, follow the instructions in the \texttt{Main script overview} section of the README.
Specifically, choose the \texttt{metamodel\_filepath} that matches the desired objective.

If the desired objective is not one of the above three metrics, then the practitioner shoudl train a new meta-model. This can be done by following the instructions in the \texttt{Training a new meta-model} section of the README. In particular, any one of the 315 metrics can be used as the objective (or any computable function of these metrics).
Once the meta-model is trained, then the above instructions can be used.

\paragraph{Final takeaways.}
Our work gives both intuition for working with rec-sys models, as well as a meta-learning pipeline that substantially lowers the level of human involvement when designing a high-performing rec-sys model.
Specifically, practitioners can use our large-scale empirical study to gain insights such as how generalizable a given algorithm is, and how ``hard'' it will be to train a high-performing model, compared to other datasets.
On the other hand, practitioners can follow the steps in our README to quickly train a high-performing model when faced with a brand new dataset.

\subsection{Additional experiments from Section \ref{sec:reczilla}} \label{app:reczilla-ablations}
\begin{figure}
     \centering
     \begin{subfigure}[b]{0.48\textwidth}
         \centering
         \includegraphics[width=\textwidth]{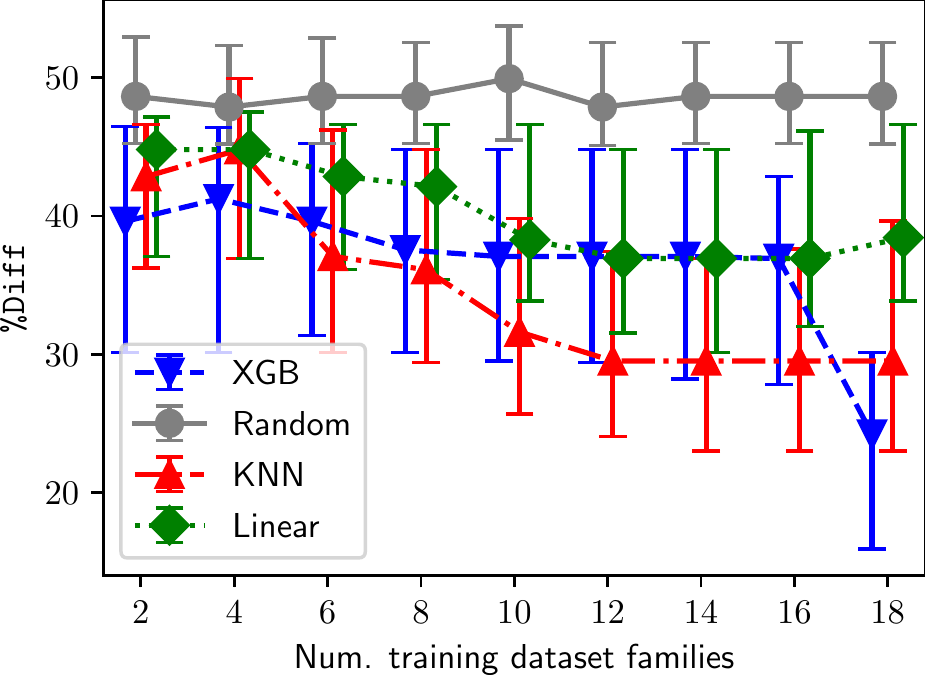}
     \end{subfigure}
     \hfill
     \begin{subfigure}[b]{0.48\textwidth}
         \centering
         \includegraphics[width=\textwidth]{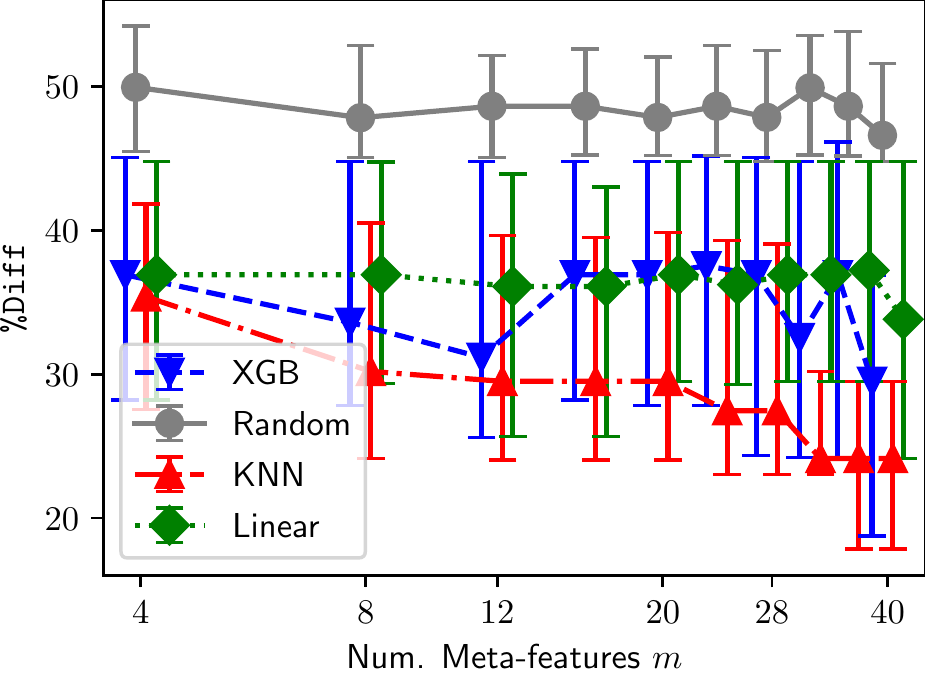}
     \end{subfigure}
        \caption{Performance of the RecZilla pipeline improves as we add more training meta-datapoints, and more meta-features $m$. Subsets of training meta-datapoints and meta-features are selected randomly, over 50 random trials. 
        Points show the median $\%\texttt{Diff}$, and error bars show the 40th and 60th percentile over all folds and random trials.
        (Left) $m=10$ meta-features, while the number of training dataset families varies. 
        (Right) All training data is used, while $m$ varies.}
        \label{fig:performance-reczilla}
\end{figure}

In this section, we give more experiments with Reczilla.
Figure \ref{fig:performance-reczilla} (left) displays \%\texttt{Diff} vs.\ the size of the meta-training set, and Figure \ref{fig:performance-reczilla} (right) displays the results of an ablation study on the number of selected meta-features $m$.
All results are averaged over all leave-one-out folds and 50 random trials.

Generally, XGBoost and KNN outperform the linear model and the random baseline, with XGBoost achieving top performance when using 10 meta-features and the maximum number of training datasets. 
Furthermore, the number of datasets in the meta-training set matters more than the meta-learning model itself. For example, the improvement of XGBoost from 4 to 10 and to 18 training datasets is larger than the difference in performance between XGBoost and KNN at 4 and 10 training datasets, respectively.
Finally, we find that the optimal number of meta-features for XGBoost and KNN peaks between 10 and 40.

\subsection{Meta-learner results with additional metrics}
\begin{figure}
     \centering
     \begin{subfigure}[b]{0.48\textwidth}
         \centering
         \includegraphics[width=\textwidth]{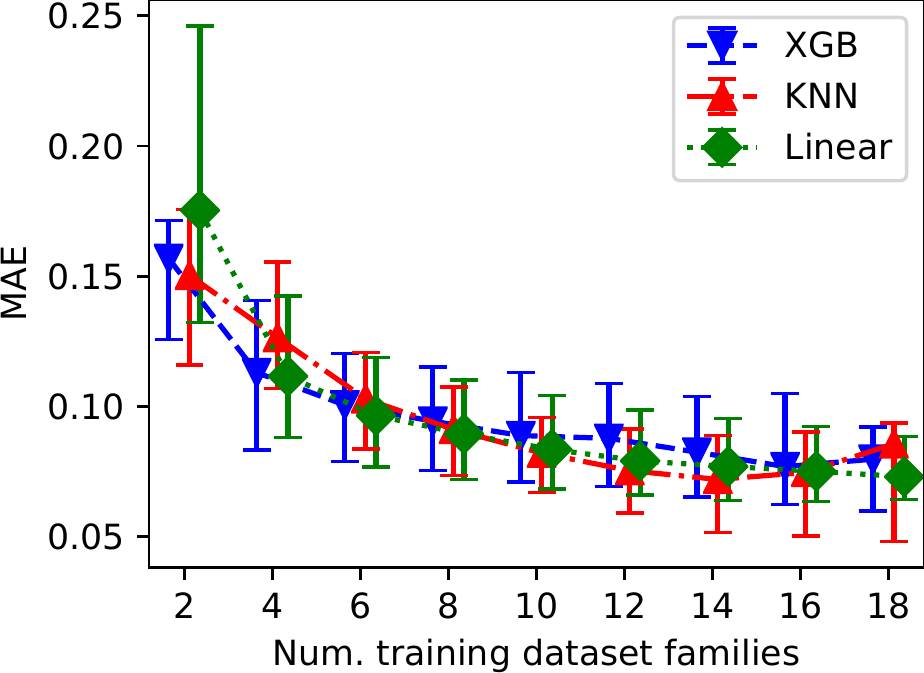}
     \end{subfigure}
     \hfill
     \begin{subfigure}[b]{0.48\textwidth}
         \centering
         \includegraphics[width=\textwidth]{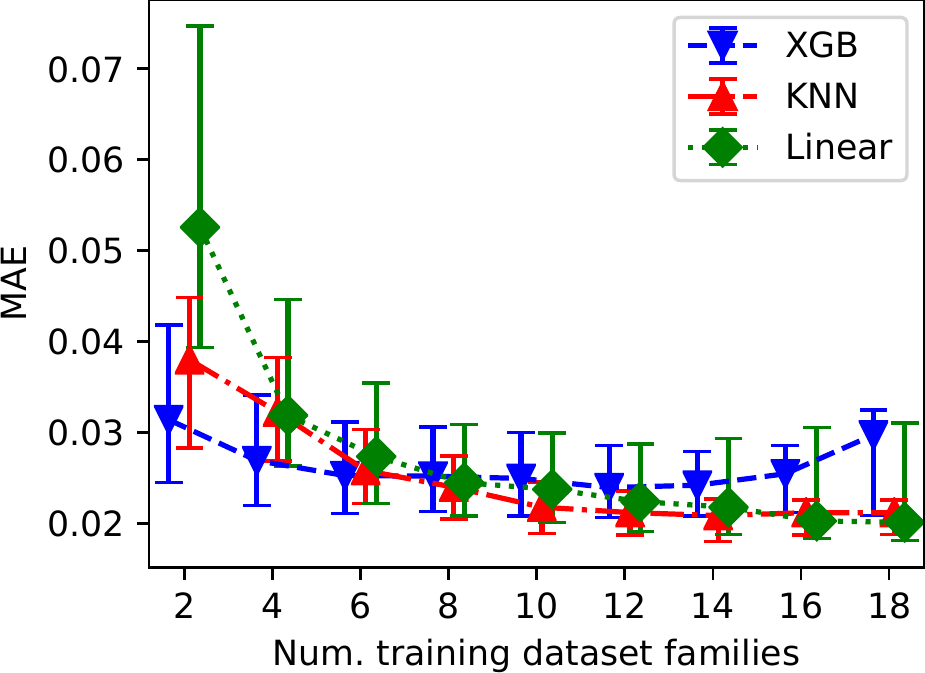}
     \end{subfigure}
     \begin{subfigure}[b]{0.48\textwidth}
         \centering
         \includegraphics[width=\textwidth]{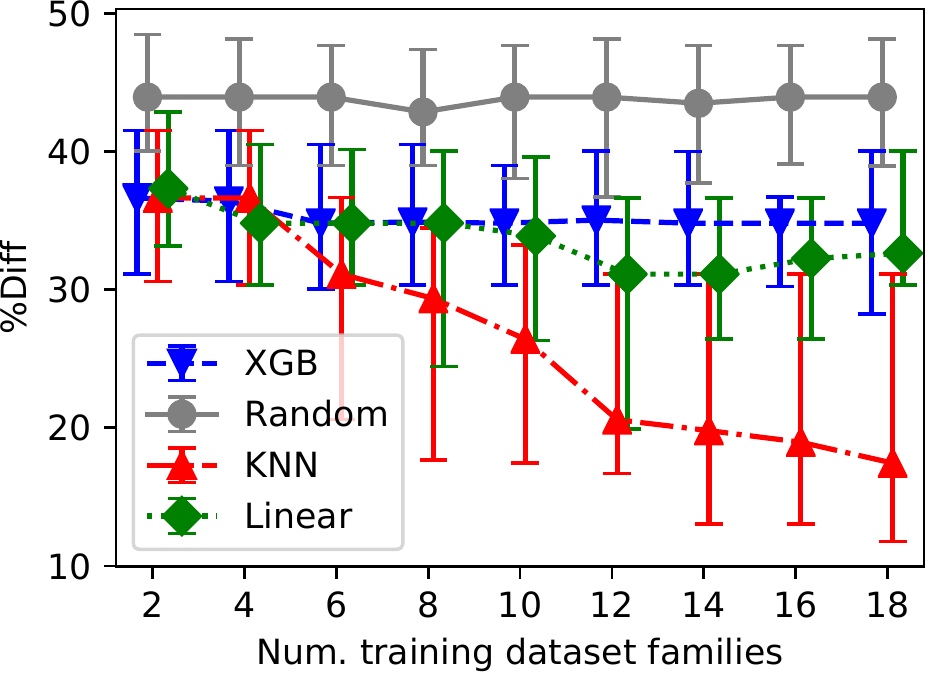}
     \end{subfigure}
     \hfill
     \begin{subfigure}[b]{0.48\textwidth}
         \centering
         \includegraphics[width=\textwidth]{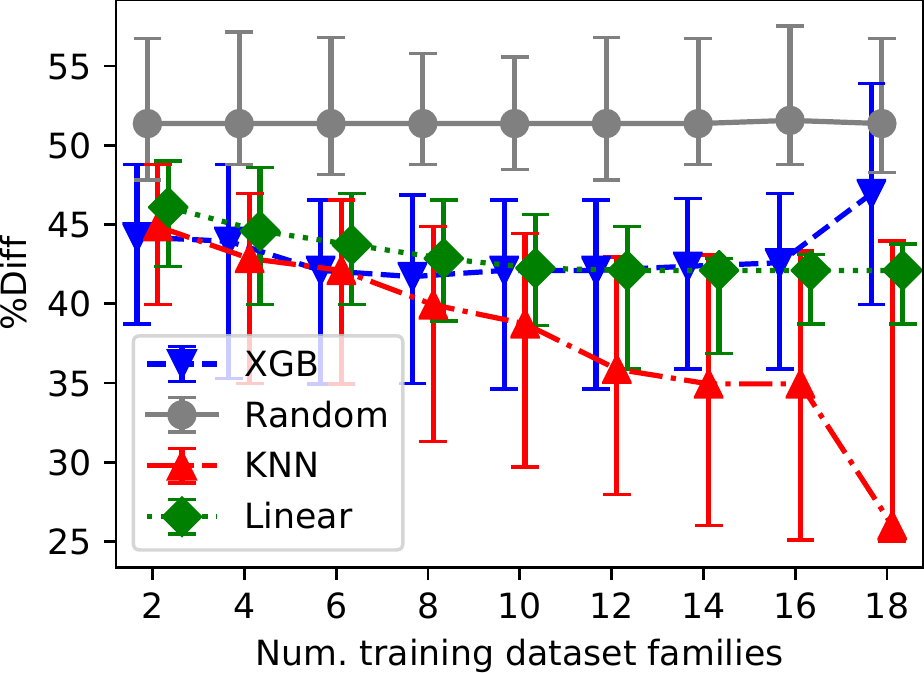}
     \end{subfigure}
     \begin{subfigure}[b]{0.48\textwidth}
         \centering
         \includegraphics[width=\textwidth]{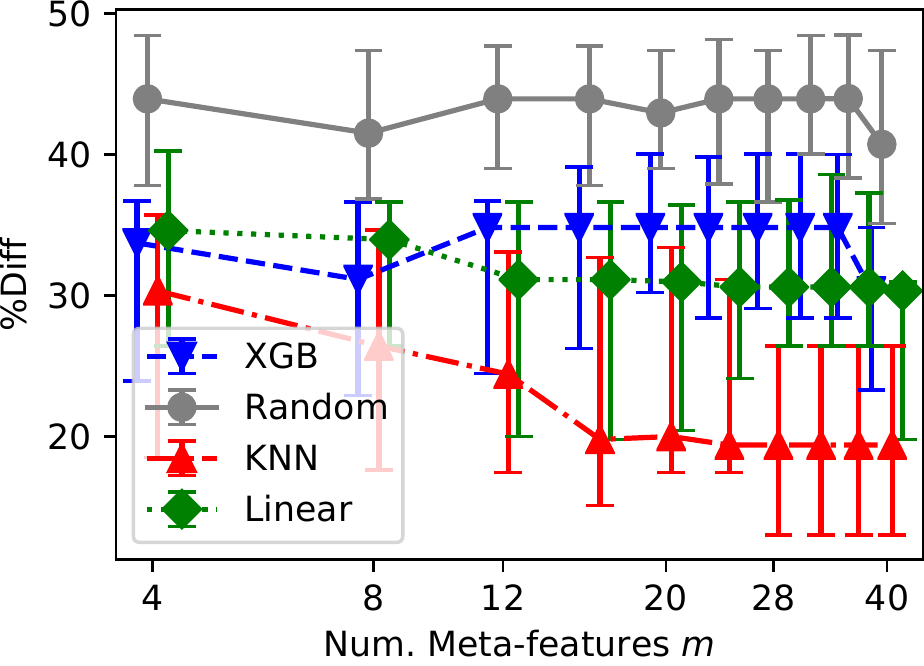}
     \end{subfigure}
     \hfill
     \begin{subfigure}[b]{0.48\textwidth}
         \centering
         \includegraphics[width=\textwidth]{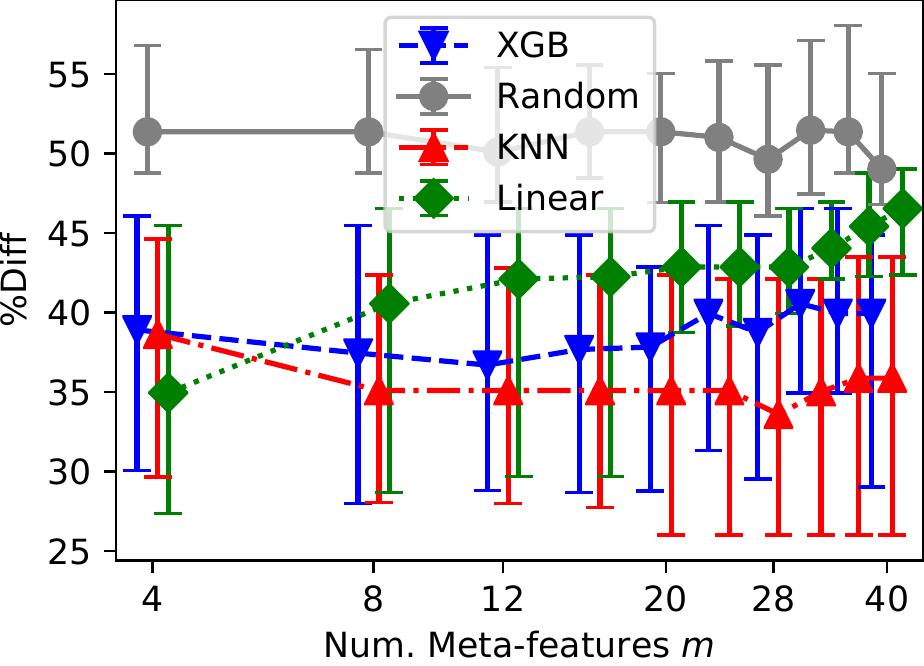}
     \end{subfigure}
        \caption{Performance of RecZilla when other base metrics are used: COVERAGE@50 (left column) and HIT-RATE@5 (right column).
        The first row shows MAE vs.\ num training dataset families (similar to Figure \ref{fig:meta-learner}),
        The second row shows \%Diff vs.\ num.\ training dataset families, and the third row shows \%Diff vs.\ num.\ meta-features (similar to Figure \ref{fig:performance-reczilla}).
        All results are averaged over all folds of leave-one-out validation and 50 trials.
        }
        \label{fig:other-metrics}
\end{figure}

The results from Figures \ref{fig:meta-learner} and \ref{fig:performance-reczilla} were generated using PREC@10 as the base recommender system metric. Now, we re-run these experiments using COVERAGE@50 and HIT-RATE@5 as the base metrics, to show variety. See Figure \ref{fig:other-metrics}.

%% file: tables/table_2a_appendix.tex
\begin{sidewaystable}
\caption{The performance of each rec-sys algorithm varies depending on the dataset and evaluation metric. Each row corresponds to a different metric (NDCG, precision, recall, and hit-rate) at a particular cutoff. Each algorithm (columns) are ranked according to each metric, values show the ``min/max (mean)'' rank over all datasets for which at least 10 algorithms produced a result. 
This table includes all 24 implemented algorithms. 
In this table, maximum rank is 24, and lower rank indicates better performance. This table is continued on the next page. (Part 1 of 3)\label{tab:all-alg-ranks}}
\begin{tabular}{lllllllll}
\toprule
{} & \rot{Item-KNN} & \rot{User-KNN} & \rot{RP3beta} & \rot{RP3beta} &  \rot{iALS} & \rot{EASE-R} & \rot{SlopeOne} & \rot{CoClustering} \\
\midrule
NDCG@1                      &      1/4 (1.5) &    2/18 (11.1) &    2/12 (5.3) &    1/12 (3.7) &  1/19 (7.4) &   1/18 (4.2) &    7/22 (15.7) &        1/20 (14.1) \\
NDCG@2                      &     1/10 (1.8) &     2/17 (9.3) &    2/14 (6.0) &    1/14 (4.0) &  1/19 (7.5) &   1/18 (4.8) &    9/23 (16.7) &        1/21 (14.9) \\
NDCG@5                      &     1/12 (2.2) &     1/16 (7.4) &    2/14 (5.9) &    1/15 (4.0) &  2/19 (7.4) &   1/18 (5.2) &   12/23 (17.8) &        2/21 (15.5) \\
NDCG@10                     &     1/13 (2.3) &     1/15 (6.6) &    2/15 (6.2) &    1/14 (4.3) &  1/19 (7.2) &   1/18 (5.2) &   14/23 (18.2) &        2/21 (15.7) \\
NDCG@50                     &     1/13 (2.9) &     1/15 (5.9) &    2/15 (6.4) &    1/16 (4.9) &  1/19 (6.5) &   1/18 (5.8) &   13/23 (18.4) &        3/21 (15.8) \\\midrule
PRECISION@1                 &      1/4 (1.5) &    2/18 (11.1) &    2/12 (5.3) &    1/12 (3.7) &  1/19 (7.4) &   1/18 (4.2) &    7/22 (15.7) &        1/20 (14.1) \\
PRECISION@2                 &     1/11 (1.9) &     2/16 (8.4) &    1/14 (5.9) &    1/14 (3.9) &  1/19 (7.4) &   1/18 (4.6) &    7/23 (16.6) &        1/21 (14.7) \\
PRECISION@5                 &     1/12 (2.6) &     1/16 (6.1) &    1/14 (5.9) &    1/13 (4.2) &  1/19 (7.1) &   1/18 (5.6) &   13/23 (17.9) &        3/21 (15.4) \\
PRECISION@10                &     1/12 (2.9) &     1/14 (4.9) &    1/14 (6.6) &    1/15 (4.6) &  1/19 (6.2) &   1/18 (6.6) &   14/23 (18.3) &        7/21 (15.9) \\
PRECISION@50                &     1/14 (4.3) &     1/12 (4.1) &    1/17 (7.3) &    1/18 (5.8) &  1/19 (5.4) &   2/18 (8.1) &   13/23 (18.1) &        2/21 (15.7) \\\midrule
RECALL@1                    &      1/4 (1.5) &    2/18 (11.1) &    2/12 (5.3) &    1/12 (3.7) &  1/19 (7.4) &   1/18 (4.2) &    7/22 (15.7) &        1/20 (14.1) \\
RECALL@2                    &     1/11 (1.9) &     2/16 (8.4) &    1/14 (5.9) &    1/14 (3.9) &  1/19 (7.4) &   1/18 (4.6) &    7/23 (16.6) &        1/21 (14.7) \\
RECALL@5                    &     1/12 (2.6) &     1/16 (6.1) &    1/14 (5.9) &    1/13 (4.2) &  1/19 (7.1) &   1/18 (5.6) &   13/23 (17.9) &        3/21 (15.4) \\
RECALL@10                   &     1/12 (2.9) &     1/14 (4.9) &    1/14 (6.6) &    1/15 (4.6) &  1/19 (6.2) &   1/18 (6.6) &   14/23 (18.3) &        7/21 (15.9) \\
RECALL@50                   &     1/14 (4.3) &     1/12 (4.1) &    1/17 (7.3) &    1/18 (5.8) &  1/19 (5.4) &   2/18 (8.1) &   13/23 (18.1) &        2/21 (15.7) \\\midrule
HIT-RATE@1                  &      1/4 (1.5) &    2/18 (11.1) &    2/12 (5.3) &    1/12 (3.7) &  1/19 (7.4) &   1/18 (4.2) &    7/22 (15.7) &        1/20 (14.1) \\
HIT-RATE@2                  &     1/11 (1.9) &     2/16 (8.4) &    1/14 (5.9) &    1/14 (3.9) &  1/19 (7.4) &   1/18 (4.6) &    7/23 (16.6) &        1/21 (14.7) \\
HIT-RATE@5                  &     1/12 (2.6) &     1/16 (6.1) &    1/14 (5.9) &    1/13 (4.2) &  1/19 (7.1) &   1/18 (5.6) &   13/23 (17.9) &        3/21 (15.4) \\
HIT-RATE@10                 &     1/12 (2.9) &     1/14 (4.9) &    1/14 (6.6) &    1/15 (4.6) &  1/19 (6.2) &   1/18 (6.6) &   14/23 (18.3) &        7/21 (15.9) \\
HIT-RATE@50                 &     1/14 (4.3) &     1/12 (4.1) &    1/17 (7.3) &    1/18 (5.8) &  1/19 (5.4) &   2/18 (8.1) &   13/23 (18.1) &        2/21 (15.7) \\
\bottomrule
\end{tabular}
\end{sidewaystable}

%% file: tables/table_2b_appendix.tex
\begin{sidewaystable}
\caption{Continuation of Table~\ref{tab:all-alg-ranks} (Part 2 of 3).\label{tab:all-alg-ranks-b}}
\begin{tabular}{llllllll}
\toprule
{} & \rot{MF-AsySVD} & \rot{MF-FunkSVD} &   \rot{SVD} &   \rot{NMF} & \rot{MF-BPR} & \rot{SLIM-BPR} & \rot{SLIM-ElasticNet} \\
\midrule
NDCG@1                      &     2/17 (10.9) &       1/13 (8.5) &  1/13 (6.1) &  1/12 (7.0) &  1/17 (10.4) &     1/12 (4.1) &            1/17 (5.8) \\
NDCG@2                      &     2/18 (11.3) &       1/17 (9.7) &  1/15 (6.0) &  2/13 (7.0) &  2/17 (11.0) &     1/13 (4.2) &            1/17 (6.2) \\
NDCG@5                      &     4/18 (11.3) &      1/15 (10.0) &  2/12 (6.3) &  1/13 (7.6) &  3/17 (12.0) &     1/14 (4.2) &            1/17 (7.1) \\
NDCG@10                     &     4/18 (11.3) &      2/15 (10.1) &  1/12 (6.4) &  1/13 (7.2) &  2/19 (12.0) &     1/14 (5.0) &            1/17 (7.1) \\
NDCG@50                     &     5/18 (11.1) &      1/15 (10.1) &  1/16 (6.4) &  1/13 (7.0) &  2/17 (12.3) &     1/14 (6.1) &            1/16 (7.8) \\\midrule
PRECISION@1                 &     2/17 (10.9) &       1/13 (8.5) &  1/13 (6.1) &  1/12 (7.0) &  1/17 (10.4) &     1/12 (4.1) &            1/17 (5.8) \\
PRECISION@2                 &     1/18 (10.8) &       2/15 (9.5) &  1/14 (5.9) &  1/12 (6.7) &  2/16 (11.0) &     1/13 (3.8) &            1/17 (6.0) \\
PRECISION@5                 &     1/18 (10.3) &       1/15 (9.3) &  1/11 (5.8) &  1/14 (7.2) &  1/17 (11.3) &     1/14 (4.7) &            1/17 (7.6) \\
PRECISION@10                &     1/18 (10.0) &       1/14 (9.4) &  1/11 (6.0) &  1/15 (6.9) &  1/18 (11.4) &     1/14 (5.3) &            1/17 (8.1) \\
PRECISION@50                &      1/16 (9.7) &       1/15 (9.0) &  1/16 (5.4) &  1/12 (6.8) &  1/17 (11.6) &     1/14 (7.3) &            1/16 (9.2) \\\midrule
RECALL@1                    &     2/17 (10.9) &       1/13 (8.5) &  1/13 (6.1) &  1/12 (7.0) &  1/17 (10.4) &     1/12 (4.1) &            1/17 (5.8) \\
RECALL@2                    &     1/18 (10.8) &       2/15 (9.5) &  1/14 (5.9) &  1/12 (6.7) &  2/16 (11.0) &     1/13 (3.8) &            1/17 (6.0) \\
RECALL@5                    &     1/18 (10.3) &       1/15 (9.3) &  1/11 (5.8) &  1/14 (7.2) &  1/17 (11.3) &     1/14 (4.7) &            1/17 (7.6) \\
RECALL@10                   &     1/18 (10.0) &       1/14 (9.4) &  1/11 (6.0) &  1/15 (6.9) &  1/18 (11.4) &     1/14 (5.3) &            1/17 (8.1) \\
RECALL@50                   &      1/16 (9.7) &       1/15 (9.0) &  1/16 (5.4) &  1/12 (6.8) &  1/17 (11.6) &     1/14 (7.3) &            1/16 (9.2) \\\midrule
HIT-RATE@1                  &     2/17 (10.9) &       1/13 (8.5) &  1/13 (6.1) &  1/12 (7.0) &  1/17 (10.4) &     1/12 (4.1) &            1/17 (5.8) \\
HIT-RATE@2                  &     1/18 (10.8) &       2/15 (9.5) &  1/14 (5.9) &  1/12 (6.7) &  2/16 (11.0) &     1/13 (3.8) &            1/17 (6.0) \\
HIT-RATE@5                  &     1/18 (10.3) &       1/15 (9.3) &  1/11 (5.8) &  1/14 (7.2) &  1/17 (11.3) &     1/14 (4.7) &            1/17 (7.6) \\
HIT-RATE@10                 &     1/18 (10.0) &       1/14 (9.4) &  1/11 (6.0) &  1/15 (6.9) &  1/18 (11.4) &     1/14 (5.3) &            1/17 (8.1) \\
HIT-RATE@50                 &      1/16 (9.7) &       1/15 (9.0) &  1/16 (5.4) &  1/12 (6.8) &  1/17 (11.6) &     1/14 (7.3) &            1/16 (9.2) \\
\bottomrule
\end{tabular}
\end{sidewaystable}

%% file: tables/table_2c_appendix.tex
\begin{sidewaystable}
\caption{Continuation of Table~\ref{tab:all-alg-ranks} (Part 3 of 3).\label{tab:all-alg-ranks-c}}
\begin{tabular}{llllllllll}
\toprule
{} & \rot{I-neural} & \rot{U-neural} & \rot{Spectral-CF} & \rot{DELF-MLP} & \rot{DELF-EF} & \rot{Mult-VAE} & \rot{GlobalEffects} & \rot{TopPop} &  \rot{Random} \\
\midrule
NDCG@1                      &   13/17 (15.0) &    1/17 (10.2) &      19/23 (21.0) &   19/24 (21.0) &  10/18 (14.0) &    4/23 (13.0) &         2/19 (12.1) &  2/17 (10.0) &  10/23 (15.9) \\
NDCG@2                      &   13/17 (15.6) &    4/19 (12.8) &      20/23 (21.6) &   20/24 (21.3) &  12/18 (14.3) &    7/20 (12.7) &         4/19 (13.4) &  3/18 (10.7) &  10/23 (16.4) \\
NDCG@5                      &   12/18 (15.8) &    2/19 (13.0) &      20/24 (22.0) &   20/24 (21.7) &   7/18 (13.7) &    8/19 (12.1) &         4/19 (13.9) &  1/19 (11.1) &  10/23 (16.6) \\
NDCG@10                     &   11/19 (15.4) &    3/18 (13.3) &      20/24 (22.4) &   20/24 (21.7) &   9/18 (14.4) &    3/19 (11.7) &         7/19 (14.0) &  1/18 (10.9) &  10/23 (16.7) \\
NDCG@50                     &   13/19 (15.8) &    1/20 (13.4) &      21/24 (23.0) &   20/24 (21.9) &  12/19 (15.1) &    1/20 (10.4) &         7/19 (14.1) &  1/18 (10.1) &   9/22 (16.6) \\\midrule
PRECISION@1                 &   13/17 (15.0) &    1/17 (10.2) &      19/23 (21.0) &   19/24 (21.0) &  10/18 (14.0) &    4/23 (13.0) &         2/19 (12.1) &  2/17 (10.0) &  10/23 (15.9) \\
PRECISION@2                 &   10/17 (15.0) &    4/19 (13.0) &      20/23 (21.6) &   20/24 (21.3) &  12/18 (14.3) &    7/19 (12.4) &         4/19 (13.4) &  1/18 (10.6) &  10/23 (16.3) \\
PRECISION@5                 &   12/19 (15.4) &    2/18 (13.1) &      20/24 (21.8) &   20/24 (21.7) &   7/18 (14.1) &    5/17 (11.4) &         2/19 (13.7) &  1/19 (10.9) &  10/22 (16.5) \\
PRECISION@10                &   12/19 (15.4) &    3/18 (13.3) &      20/24 (21.8) &   20/24 (21.4) &  11/17 (14.9) &    2/18 (10.8) &         7/19 (14.1) &  1/18 (10.7) &  10/22 (16.6) \\
PRECISION@50                &   13/19 (16.2) &    1/19 (12.9) &      22/24 (23.2) &   19/24 (21.6) &  13/19 (15.4) &     1/19 (8.6) &         7/20 (14.1) &   1/18 (9.3) &   9/22 (16.5) \\\midrule
RECALL@1                    &   13/17 (15.0) &    1/17 (10.2) &      19/23 (21.0) &   19/24 (21.0) &  10/18 (14.0) &    4/23 (13.0) &         2/19 (12.1) &  2/17 (10.0) &  10/23 (15.9) \\
RECALL@2                    &   10/17 (15.0) &    4/19 (13.0) &      20/23 (21.6) &   20/24 (21.3) &  12/18 (14.3) &    7/19 (12.4) &         4/19 (13.4) &  1/18 (10.6) &  10/23 (16.3) \\
RECALL@5                    &   12/19 (15.4) &    2/18 (13.1) &      20/24 (21.8) &   20/24 (21.7) &   7/18 (14.1) &    5/17 (11.4) &         2/19 (13.7) &  1/19 (10.9) &  10/22 (16.5) \\
RECALL@10                   &   12/19 (15.4) &    3/18 (13.3) &      20/24 (21.8) &   20/24 (21.4) &  11/17 (14.9) &    2/18 (10.8) &         7/19 (14.1) &  1/18 (10.7) &  10/22 (16.6) \\
RECALL@50                   &   13/19 (16.2) &    1/19 (12.9) &      22/24 (23.2) &   19/24 (21.6) &  13/19 (15.4) &     1/19 (8.6) &         7/20 (14.1) &   1/18 (9.3) &   9/22 (16.5) \\\midrule
HIT-RATE@1                  &   13/17 (15.0) &    1/17 (10.2) &      19/23 (21.0) &   19/24 (21.0) &  10/18 (14.0) &    4/23 (13.0) &         2/19 (12.1) &  2/17 (10.0) &  10/23 (15.9) \\
HIT-RATE@2                  &   10/17 (15.0) &    4/19 (13.0) &      20/23 (21.6) &   20/24 (21.3) &  12/18 (14.3) &    7/19 (12.4) &         4/19 (13.4) &  1/18 (10.6) &  10/23 (16.3) \\
HIT-RATE@5                  &   12/19 (15.4) &    2/18 (13.1) &      20/24 (21.8) &   20/24 (21.7) &   7/18 (14.1) &    5/17 (11.4) &         2/19 (13.7) &  1/19 (10.9) &  10/22 (16.5) \\
HIT-RATE@10                 &   12/19 (15.4) &    3/18 (13.3) &      20/24 (21.8) &   20/24 (21.4) &  11/17 (14.9) &    2/18 (10.8) &         7/19 (14.1) &  1/18 (10.7) &  10/22 (16.6) \\
HIT-RATE@50                 &   13/19 (16.2) &    1/19 (12.9) &      22/24 (23.2) &   19/24 (21.6) &  13/19 (15.4) &     1/19 (8.6) &         7/20 (14.1) &   1/18 (9.3) &   9/22 (16.5) \\
\bottomrule
\end{tabular}

\end{sidewaystable}

%% file: main.bbl
\begin{thebibliography}{10}

\bibitem{adeniyi2016automated}
David~Adedayo Adeniyi, Zhaoqiang Wei, and Y~Yongquan.
\newblock Automated web usage data mining and recommendation system using
  k-nearest neighbor (knn) classification method.
\newblock {\em Applied Computing and Informatics}, 12(1):90--108, 2016.

\bibitem{adomavicius2011improving}
Gediminas Adomavicius and YoungOk Kwon.
\newblock Improving aggregate recommendation diversity using ranking-based
  techniques.
\newblock {\em IEEE Transactions on Knowledge and Data Engineering},
  24(5):896--911, 2011.

\bibitem{adomavicius2012impact}
Gediminas Adomavicius and Jingjing Zhang.
\newblock Impact of data characteristics on recommender systems performance.
\newblock {\em ACM Transactions on Management Information Systems (TMIS)},
  3(1):1--17, 2012.

\bibitem{aggarwal2016recommender}
Charu~C Aggarwal et~al.
\newblock {\em Recommender systems}, volume~1.
\newblock Springer, 2016.

\bibitem{asymcos}
Fabio Aiolli.
\newblock Efficient top-n recommendation for very large scale binary rated
  datasets.
\newblock In {\em Proceedings of the 7th ACM Conference on Recommender
  Systems}, RecSys '13, page 273–280, New York, NY, USA, 2013. Association
  for Computing Machinery.

\bibitem{auto-surprise}
Rohan Anand and Joeran Beel.
\newblock Auto-surprise: An automated recommender-system (autorecsys) library
  with tree of parzens estimator (tpe) optimization.
\newblock In {\em Fourteenth ACM Conference on Recommender Systems}, pages
  585--587, 2020.

\bibitem{slim-bpr}
Krisztian Balog, Filip Radlinski, and Shushan Arakelyan.
\newblock Transparent, scrutable and explainable user models for personalized
  recommendation.
\newblock In {\em Proceedings of the 42nd International ACM SIGIR Conference on
  Research and Development in Information Retrieval}, SIGIR'19, page 265–274,
  New York, NY, USA, 2019. Association for Computing Machinery.

\bibitem{frappe15}
Linas Baltrunas, Karen Church, Alexandros Karatzoglou, and Nuria Oliver.
\newblock Frappe: Understanding the usage and perception of mobile app
  recommendations in-the-wild.
\newblock {\em arXiv preprint arXiv:1505.03014}, 2015.

\bibitem{netflixprize}
James Bennett, Stan Lanning, et~al.
\newblock The netflix prize.
\newblock In {\em Proceedings of KDD cup and workshop}, volume 2007, page~35.
  Citeseer, 2007.

\bibitem{BOBADILLA2013109}
J.~Bobadilla, F.~Ortega, A.~Hernando, and A.~Gutiérrez.
\newblock Recommender systems survey.
\newblock {\em Knowledge-Based Systems}, 46:109--132, 2013.

\bibitem{Cantador:RecSys2011}
Iv\'{a}n Cantador, Peter Brusilovsky, and Tsvi Kuflik.
\newblock 2nd workshop on information heterogeneity and fusion in recommender
  systems (hetrec 2011).
\newblock In {\em Proceedings of the 5th ACM conference on Recommender
  systems}, RecSys 2011, New York, NY, USA, 2011. ACM.

\bibitem{chen2016xgboost}
Tianqi Chen and Carlos Guestrin.
\newblock Xgboost: A scalable tree boosting system.
\newblock In {\em Proceedings of the 22nd acm sigkdd international conference
  on knowledge discovery and data mining}, pages 785--794, 2016.

\bibitem{ijcai2018-462}
Weiyu Cheng, Yanyan Shen, Yanmin Zhu, and Linpeng Huang.
\newblock Delf: A dual-embedding based deep latent factor model for
  recommendation.
\newblock In {\em Proceedings of the Twenty-Seventh International Joint
  Conference on Artificial Intelligence, {IJCAI-18}}, pages 3329--3335.
  International Joint Conferences on Artificial Intelligence Organization, 7
  2018.

\bibitem{chin2022datasets}
Jin~Yao Chin, Yile Chen, and Gao Cong.
\newblock The datasets dilemma: How much do we really know about recommendation
  datasets?
\newblock In {\em Proceedings of the Fifteenth ACM International Conference on
  Web Search and Data Mining}, pages 141--149, 2022.

\bibitem{gowalla}
Eunjoon Cho, Seth~A. Myers, and Jure Leskovec.
\newblock Friendship and mobility: User movement in location-based social
  networks.
\newblock In {\em Proceedings of the 17th ACM SIGKDD International Conference
  on Knowledge Discovery and Data Mining}, KDD '11, page 1082–1090, New York,
  NY, USA, 2011. Association for Computing Machinery.

\bibitem{cho2018analyzing}
Hyerim Cho, Marc~L Schmalz, Stephen~A Keating, and Jin~Ha Lee.
\newblock Analyzing anime users’ online forum queries for recommendation
  using content analysis.
\newblock {\em Journal of Documentation}, 2018.

\bibitem{collins2019meta}
Andrew Collins and Joeran Beel.
\newblock Meta-learned per-instance algorithm selection in scholarly
  recommender systems.
\newblock {\em arXiv preprint arXiv:1912.08694}, 2019.

\bibitem{collins2018novel}
Andrew Collins, Dominika Tkaczyk, and Joeran Beel.
\newblock A novel approach to recommendation algorithm selection using
  meta-learning.
\newblock In {\em AICS}, pages 210--219, 2018.

\bibitem{collins2018one}
Andrew Collins, Dominika Tkaczyk, and Joeran Beel.
\newblock One-at-a-time: A meta-learning recommender-system for
  recommendation-algorithm selection on micro level.
\newblock {\em arXiv preprint arXiv:1805.12118}, 2018.

\bibitem{cooper2014random}
Colin Cooper, Sang~Hyuk Lee, Tomasz Radzik, and Yiannis Siantos.
\newblock Random walks in recommender systems: exact computation and
  simulations.
\newblock In {\em Proceedings of the 23rd international conference on world
  wide web}, pages 811--816, 2014.

\bibitem{covington2016deep}
Paul Covington, Jay Adams, and Emre Sargin.
\newblock Deep neural networks for youtube recommendations.
\newblock In {\em Proceedings of the 10th ACM conference on recommender
  systems}, pages 191--198, 2016.

\bibitem{cremonesi2010performance}
Paolo Cremonesi, Yehuda Koren, and Roberto Turrin.
\newblock Performance of recommender algorithms on top-n recommendation tasks.
\newblock In {\em Proceedings of the fourth ACM conference on Recommender
  systems}, pages 39--46, 2010.

\bibitem{cunha2016selecting}
Tiago Cunha, Carlos Soares, and Andr{\'e}~CPLF de~Carvalho.
\newblock Selecting collaborative filtering algorithms using metalearning.
\newblock In {\em Joint European Conference on Machine Learning and Knowledge
  Discovery in Databases}, pages 393--409. Springer, 2016.

\bibitem{cunha2017recommending}
Tiago Cunha, Carlos Soares, and Andr{\'e}~CPLF de~Carvalho.
\newblock Recommending collaborative filtering algorithms using subsampling
  landmarkers.
\newblock In {\em International Conference on Discovery Science}, pages
  189--203. Springer, 2017.

\bibitem{cunha2018cf4cfmeta}
Tiago Cunha, Carlos Soares, and Andr{\'e}~CPLF de~Carvalho.
\newblock Cf4cf-meta: Hybrid collaborative filtering algorithm selection
  framework.
\newblock In {\em International Conference on Discovery Science}, pages
  114--128. Springer, 2018.

\bibitem{cunha2018cf4cf}
Tiago Cunha, Carlos Soares, and Andr{\'e}~CPLF de~Carvalho.
\newblock Cf4cf: recommending collaborative filtering algorithms using
  collaborative filtering.
\newblock In {\em Proceedings of the 12th ACM Conference on Recommender
  Systems}, pages 357--361, 2018.

\bibitem{cunha2018metalearning}
Tiago Cunha, Carlos Soares, and Andr{\'e}~CPLF de~Carvalho.
\newblock Metalearning and recommender systems: A literature review and
  empirical study on the algorithm selection problem for collaborative
  filtering.
\newblock {\em Information Sciences}, 423:128--144, 2018.

\bibitem{dacrema2021troubling}
Maurizio~Ferrari Dacrema, Simone Boglio, Paolo Cremonesi, and Dietmar Jannach.
\newblock A troubling analysis of reproducibility and progress in recommender
  systems research.
\newblock {\em ACM Transactions on Information Systems (TOIS)}, 39(2):1--49,
  2021.

\bibitem{DacremaCJ19}
Maurizio~Ferrari Dacrema, Paolo Cremonesi, and Dietmar Jannach.
\newblock Are we really making much progress? {A} worrying analysis of recent
  neural recommendation approaches.
\newblock In Toine Bogers, Alan Said, Peter Brusilovsky, and Domonkos Tikk,
  editors, {\em Proceedings of the 13th {ACM} Conference on Recommender
  Systems, RecSys 2019, Copenhagen, Denmark, September 16-20, 2019}, pages
  101--109. {ACM}, 2019.

\bibitem{deng2009imagenet}
Jia Deng, Wei Dong, Richard Socher, Li-Jia Li, Kai Li, and Li~Fei-Fei.
\newblock Imagenet: A large-scale hierarchical image database.
\newblock In {\em 2009 IEEE conference on computer vision and pattern
  recognition}, pages 248--255. Ieee, 2009.

\bibitem{bert}
Jacob Devlin, Ming-Wei Chang, Kenton Lee, and Kristina Toutanova.
\newblock {BERT}: Pre-training of deep bidirectional transformers for language
  understanding.
\newblock In {\em Proceedings of the 2019 Conference of the North {A}merican
  Chapter of the Association for Computational Linguistics: Human Language
  Technologies, Volume 1 (Long and Short Papers)}, pages 4171--4186,
  Minneapolis, Minnesota, June 2019. Association for Computational Linguistics.

\bibitem{dice}
Lee~R. Dice.
\newblock Measures of the amount of ecologic association between species.
\newblock {\em Ecology}, 26(3):297--302, 1945.

\bibitem{Dooms13crowdrec}
Simon Dooms, Toon De~Pessemier, and Luc Martens.
\newblock Movietweetings: a movie rating dataset collected from twitter.
\newblock In {\em Workshop on Crowdsourcing and Human Computation for
  Recommender Systems, CrowdRec at RecSys 2013}, 2013.

\bibitem{dror2012yahoo}
Gideon Dror, Noam Koenigstein, Yehuda Koren, and Markus Weimer.
\newblock The yahoo! music dataset and kdd-cup’11.
\newblock In {\em Proceedings of KDD Cup 2011}, pages 3--18. PMLR, 2012.

\bibitem{ekstrand2012recommenders}
Michael Ekstrand and John Riedl.
\newblock When recommenders fail: predicting recommender failure for algorithm
  selection and combination.
\newblock In {\em Proceedings of the sixth ACM conference on Recommender
  systems}, pages 233--236, 2012.

\bibitem{nas-survey}
Thomas Elsken, Jan~Hendrik Metzen, and Frank Hutter.
\newblock Neural architecture search: A survey.
\newblock In {\em JMLR}, 2019.

\bibitem{frankowski-wikilens}
Dan Frankowski, Shyong~K. Lam, Shilad Sen, F.~Maxwell Harper, Scott Yilek,
  Michael Cassano, and John Riedl.
\newblock Recommenders everywhere: The wikilens community-maintained
  recommender system.
\newblock In {\em Proceedings of the 2007 International Symposium on Wikis},
  WikiSym '07, page 47–60, New York, NY, USA, 2007. Association for Computing
  Machinery.

\bibitem{ge2020understanding}
Yingqiang Ge, Shuya Zhao, Honglu Zhou, Changhua Pei, Fei Sun, Wenwu Ou, and
  Yongfeng Zhang.
\newblock Understanding echo chambers in e-commerce recommender systems.
\newblock In {\em Proceedings of the 43rd international ACM SIGIR conference on
  research and development in information retrieval}, pages 2261--2270, 2020.

\bibitem{george2005scalable}
Thomas George and Srujana Merugu.
\newblock A scalable collaborative filtering framework based on co-clustering.
\newblock In {\em Fifth IEEE International Conference on Data Mining
  (ICDM'05)}, pages 4--pp. IEEE, 2005.

\bibitem{jester2}
Ken Goldberg, Theresa Roeder, Dhruv Gupta, and Chris Perkins.
\newblock Eigentaste: A constant time collaborative filtering algorithm.
\newblock {\em Information Retrieval}, 4(2):133--151, 2001.

\bibitem{gomez2015netflix}
Carlos~A Gomez-Uribe and Neil Hunt.
\newblock The netflix recommender system: Algorithms, business value, and
  innovation.
\newblock {\em ACM Transactions on Management Information Systems (TMIS)},
  6(4):1--19, 2015.

\bibitem{gravino2019towards}
Pietro Gravino, Bernardo Monechi, and Vittorio Loreto.
\newblock Towards novelty-driven recommender systems.
\newblock {\em Comptes Rendus Physique}, 20(4):371--379, 2019.

\bibitem{griffith2012investigations}
Josephine Griffith, Colm O'Riordan, and Humphrey Sorensen.
\newblock Investigations into user rating information and predictive accuracy
  in a collaborative filtering domain.
\newblock In {\em Proceedings of the 27th Annual ACM Symposium on Applied
  Computing}, pages 937--942, 2012.

\bibitem{guo2013novel}
G.~Guo, J.~Zhang, and N.~Yorke-Smith.
\newblock A novel bayesian similarity measure for recommender systems.
\newblock In {\em Proceedings of the 23rd International Joint Conference on
  Artificial Intelligence (IJCAI)}, pages 2619--2625, 2013.

\bibitem{etaf-ciaodvd}
Guibing Guo, Jie Zhang, Daniel Thalmann, and Neil Yorke-Smith.
\newblock Etaf: An extended trust antecedents framework for trust prediction.
\newblock In {\em 2014 IEEE/ACM International Conference on Advances in Social
  Networks Analysis and Mining (ASONAM 2014)}, pages 540--547, 2014.

\bibitem{gupta2021enpso}
Garima Gupta and Rahul Katarya.
\newblock Enpso: An automl technique for generating ensemble recommender
  system.
\newblock {\em Arabian Journal for Science and Engineering}, 46(9):8677--8695,
  2021.

\bibitem{gupta2020auto}
Srijan Gupta.
\newblock Auto-caserec: A novel automated recommender system framework, 2020.

\bibitem{haldar2019applying}
Malay Haldar, Mustafa Abdool, Prashant Ramanathan, Tao Xu, Shulin Yang,
  Huizhong Duan, Qing Zhang, Nick Barrow-Williams, Bradley~C Turnbull,
  Brendan~M Collins, et~al.
\newblock Applying deep learning to airbnb search.
\newblock In {\em Proceedings of the 25th ACM SIGKDD International Conference
  on Knowledge Discovery \& Data Mining}, pages 1927--1935, 2019.

\bibitem{movielens}
F.~Maxwell Harper and Joseph~A. Konstan.
\newblock The movielens datasets: History and context.
\newblock {\em ACM Trans. Interact. Intell. Syst.}, 5(4), dec 2015.

\bibitem{he2019rethinking}
Kaiming He, Ross Girshick, and Piotr Doll{\'a}r.
\newblock Rethinking imagenet pre-training.
\newblock In {\em Proceedings of the IEEE/CVF International Conference on
  Computer Vision}, pages 4918--4927, 2019.

\bibitem{hu2008collaborative}
Yifan Hu, Yehuda Koren, and Chris Volinsky.
\newblock Collaborative filtering for implicit feedback datasets.
\newblock In {\em 2008 Eighth IEEE international conference on data mining},
  pages 263--272. Ieee, 2008.

\bibitem{huang2011does}
Zan Huang and Daniel~Dajun Zeng.
\newblock Why does collaborative filtering work? transaction-based
  recommendation model validation and selection by analyzing bipartite random
  graphs.
\newblock {\em INFORMS Journal on Computing}, 23(1):138--152, 2011.

\bibitem{Hug2020}
Nicolas Hug.
\newblock Surprise: A python library for recommender systems.
\newblock {\em Journal of Open Source Software}, 5(52):2174, 2020.

\bibitem{jannach2010recommender}
Dietmar Jannach, Markus Zanker, Alexander Felfernig, and Gerhard Friedrich.
\newblock {\em Recommender systems: an introduction}.
\newblock Cambridge University Press, 2010.

\bibitem{jiang2019degenerate}
Ray Jiang, Silvia Chiappa, Tor Lattimore, Andr{\'a}s Gy{\"o}rgy, and Pushmeet
  Kohli.
\newblock Degenerate feedback loops in recommender systems.
\newblock In {\em Proceedings of the 2019 AAAI/ACM Conference on AI, Ethics,
  and Society}, pages 383--390, 2019.

\bibitem{asysvd}
Yehuda Koren.
\newblock Factorization meets the neighborhood: A multifaceted collaborative
  filtering model.
\newblock In {\em Proceedings of the 14th ACM SIGKDD International Conference
  on Knowledge Discovery and Data Mining}, KDD '08, page 426–434, New York,
  NY, USA, 2008. Association for Computing Machinery.

\bibitem{koren2009matrix}
Yehuda Koren, Robert Bell, and Chris Volinsky.
\newblock Matrix factorization techniques for recommender systems.
\newblock {\em Computer}, 42(8):30--37, 2009.

\bibitem{kouki2020lab}
Pigi Kouki, Ilias Fountalis, Nikolaos Vasiloglou, Xiquan Cui, Edo Liberty, and
  Khalifeh Al~Jadda.
\newblock From the lab to production: A case study of session-based
  recommendations in the home-improvement domain.
\newblock In {\em Fourteenth ACM conference on recommender systems}, pages
  140--149, 2020.

\bibitem{konect}
J\'{e}r\^{o}me Kunegis.
\newblock Konect: The koblenz network collection.
\newblock In {\em Proceedings of the 22nd International Conference on World
  Wide Web}, WWW '13 Companion, page 1343–1350, New York, NY, USA, 2013.
  Association for Computing Machinery.

\bibitem{online-dating}
J\'{e}r\^{o}me Kunegis, Gerd Gr\"{o}ner, and Thomas Gottron.
\newblock Online dating recommender systems: The split-complex number approach.
\newblock In {\em Proceedings of the 4th ACM RecSys Workshop on Recommender
  Systems and the Social Web}, RSWeb '12, page 37–44, New York, NY, USA,
  2012. Association for Computing Machinery.

\bibitem{lemire2005slope}
Daniel Lemire and Anna Maclachlan.
\newblock Slope one predictors for online rating-based collaborative filtering.
\newblock In {\em Proceedings of the 2005 SIAM International Conference on Data
  Mining}, pages 471--475. SIAM, 2005.

\bibitem{levy2013efficient}
Mark Levy and Kris Jack.
\newblock Efficient top-n recommendation by linear regression, 2013.

\bibitem{levy2014neural}
Omer Levy and Yoav Goldberg.
\newblock Neural word embedding as implicit matrix factorization.
\newblock {\em Advances in neural information processing systems}, 27, 2014.

\bibitem{multi-vae}
Dawen Liang, Rahul~G. Krishnan, Matthew~D. Hoffman, and Tony Jebara.
\newblock Variational autoencoders for collaborative filtering.
\newblock In {\em Proceedings of the 2018 World Wide Web Conference}, WWW '18,
  page 689–698, Republic and Canton of Geneva, CHE, 2018. International World
  Wide Web Conferences Steering Committee.

\bibitem{liu2019roberta}
Yinhan Liu, Myle Ott, Naman Goyal, Jingfei Du, Mandar Joshi, Danqi Chen, Omer
  Levy, Mike Lewis, Luke Zettlemoyer, and Veselin Stoyanov.
\newblock Roberta: A robustly optimized bert pretraining approach, 2019.

\bibitem{majumder-etal-2019-generating}
Bodhisattwa~Prasad Majumder, Shuyang Li, Jianmo Ni, and Julian McAuley.
\newblock Generating personalized recipes from historical user preferences.
\newblock In {\em Proceedings of the 2019 Conference on Empirical Methods in
  Natural Language Processing and the 9th International Joint Conference on
  Natural Language Processing (EMNLP-IJCNLP)}, pages 5976--5982, Hong Kong,
  China, November 2019. Association for Computational Linguistics.

\bibitem{trust-aware-recsys}
Paolo Massa and Paolo Avesani.
\newblock Trust-aware recommender systems.
\newblock In {\em Proceedings of the 2007 ACM Conference on Recommender
  Systems}, RecSys '07, page 17–24, New York, NY, USA, 2007. Association for
  Computing Machinery.

\bibitem{massa2008trustlet}
Paolo Massa, Kasper Souren, Martino Salvetti, and Danilo Tomasoni.
\newblock Trustlet, open research on trust metrics.
\newblock {\em Scalable Computing: Practice and Experience}, 9(4), 2008.

\bibitem{morini2021toward}
Virginia Morini, Laura Pollacci, and Giulio Rossetti.
\newblock Toward a standard approach for echo chamber detection: Reddit case
  study.
\newblock {\em Applied Sciences}, 11(12):5390, 2021.

\bibitem{murphyjaccard}
Allan~H. Murphy.
\newblock The finley affair: A signal event in the history of forecast
  verification.
\newblock {\em Weather and Forecasting}, 11(1):3 -- 20, 1996.

\bibitem{ni-etal-2019-justifying}
Jianmo Ni, Jiacheng Li, and Julian McAuley.
\newblock Justifying recommendations using distantly-labeled reviews and
  fine-grained aspects.
\newblock In {\em Proceedings of the 2019 Conference on Empirical Methods in
  Natural Language Processing and the 9th International Joint Conference on
  Natural Language Processing (EMNLP-IJCNLP)}, pages 188--197, Hong Kong,
  China, November 2019. Association for Computational Linguistics.

\bibitem{paudel2016updatable}
Bibek Paudel, Fabian Christoffel, Chris Newell, and Abraham Bernstein.
\newblock Updatable, accurate, diverse, and scalable recommendations for
  interactive applications.
\newblock {\em ACM Transactions on Interactive Intelligent Systems (TiiS)},
  7(1):1--34, 2016.

\bibitem{rendle2012bpr}
Steffen Rendle, Christoph Freudenthaler, Zeno Gantner, and Lars Schmidt-Thieme.
\newblock {BPR}: Bayesian personalized ranking from implicit feedback.
\newblock In {\em Proceedings of the Twenty-Fifth Conference on Uncertainty in
  Artificial Intelligence}, UAI '09, page 452–461, Arlington, Virginia, USA,
  2009. AUAI Press.

\bibitem{userknn}
Paul Resnick, Neophytos Iacovou, Mitesh Suchak, Peter Bergstrom, and John
  Riedl.
\newblock Grouplens: An open architecture for collaborative filtering of
  netnews.
\newblock In {\em Proceedings of the 1994 ACM Conference on Computer Supported
  Cooperative Work}, CSCW '94, page 175–186, New York, NY, USA, 1994.
  Association for Computing Machinery.

\bibitem{ridnik2021imagenet}
Tal Ridnik, Emanuel Ben-Baruch, Asaf Noy, and Lihi Zelnik.
\newblock Imagenet-21k pretraining for the masses.
\newblock In J.~Vanschoren and S.~Yeung, editors, {\em Proceedings of the
  Neural Information Processing Systems Track on Datasets and Benchmarks},
  volume~1, 2021.

\bibitem{itemknn}
Badrul Sarwar, George Karypis, Joseph Konstan, and John Riedl.
\newblock Item-based collaborative filtering recommendation algorithms.
\newblock In {\em Proceedings of the 10th International Conference on World
  Wide Web}, WWW '01, page 285–295, New York, NY, USA, 2001. Association for
  Computing Machinery.

\bibitem{smith2017two}
Brent Smith and Greg Linden.
\newblock Two decades of recommender systems at amazon. com.
\newblock {\em Ieee internet computing}, 21(3):12--18, 2017.

\bibitem{steck2019embarrassingly}
Harald Steck.
\newblock Embarrassingly shallow autoencoders for sparse data.
\newblock In {\em The World Wide Web Conference}, pages 3251--3257, 2019.

\bibitem{tversky1977features}
Amos Tversky.
\newblock Features of similarity.
\newblock {\em Psychological review}, 84(4):327, 1977.

\bibitem{wan-marketbias}
Mengting Wan, Jianmo Ni, Rishabh Misra, and Julian McAuley.
\newblock {\em Addressing Marketing Bias in Product Recommendations}, page
  618–626.
\newblock Association for Computing Machinery, New York, NY, USA, 2020.

\bibitem{wang2019sequential}
Fan Wang, Xiaomin Fang, Lihang Liu, Yaxue Chen, Jiucheng Tao, Zhiming Peng,
  Cihang Jin, and Hao Tian.
\newblock Sequential evaluation and generation framework for combinatorial
  recommender system.
\newblock {\em arXiv preprint arXiv:1902.00245}, 2019.

\bibitem{auto-rec}
Ting-Hsiang Wang, Xia Hu, Haifeng Jin, Qingquan Song, Xiaotian Han, and Zirui
  Liu.
\newblock {\em AutoRec: An Automated Recommender System}, page 582–584.
\newblock Association for Computing Machinery, New York, NY, USA, 2020.

\bibitem{satzilla}
Lin Xu, Frank Hutter, Holger~H. Hoos, and Kevin Leyton-Brown.
\newblock {SATzilla:} portfolio-based algorithm selection for {SAT}.
\newblock {\em Journal of Artificial Intelligence Research}, 32:565--606, June
  2008.

\bibitem{neurec}
Shuai Zhang, Lina Yao, Aixin Sun, Sen Wang, Guodong Long, and Manqing Dong.
\newblock Neurec: On nonlinear transformation for personalized ranking.
\newblock In {\em Proceedings of the Twenty-Seventh International Joint
  Conference on Artificial Intelligence, {IJCAI-18}}, pages 3669--3675.
  International Joint Conferences on Artificial Intelligence Organization, 7
  2018.

\bibitem{spectral-cf}
Lei Zheng, Chun-Ta Lu, Fei Jiang, Jiawei Zhang, and Philip~S. Yu.
\newblock Spectral collaborative filtering.
\newblock In {\em Proceedings of the 12th ACM Conference on Recommender
  Systems}, RecSys '18, page 311–319, New York, NY, USA, 2018. Association
  for Computing Machinery.

\bibitem{zhou2010solving}
Tao Zhou, Zolt{\'a}n Kuscsik, Jian-Guo Liu, Mat{\'u}{\v{s}} Medo,
  Joseph~Rushton Wakeling, and Yi-Cheng Zhang.
\newblock Solving the apparent diversity-accuracy dilemma of recommender
  systems.
\newblock {\em Proceedings of the National Academy of Sciences},
  107(10):4511--4515, 2010.

\bibitem{book-crossing}
Cai-Nicolas Ziegler, Sean~M. McNee, Joseph~A. Konstan, and Georg Lausen.
\newblock Improving recommendation lists through topic diversification.
\newblock In {\em Proceedings of the 14th International Conference on World
  Wide Web}, WWW '05, page 22–32, New York, NY, USA, 2005. Association for
  Computing Machinery.

\end{thebibliography}
